\documentclass[aps,prc,preprintnumbers,superscriptaddress,floatfix,showpacs]{revtex4-1}
\usepackage{amsmath}

\usepackage{epsfig}
\usepackage{dcolumn}
\usepackage{bm}
\usepackage{amssymb}
\usepackage{xcolor}

\begin{document}

\title{Bulk Viscosity Effects in Event-by-Event Relativistic Hydrodynamics}

\author{Jacquelyn Noronha-Hostler}
\affiliation{Instituto de F\'{i}sica, Universidade de S\~{a}o Paulo, C.P.
66318, 05315-970 S\~{a}o Paulo, SP, Brazil}

\author{Gabriel S.~Denicol}
\affiliation{Department of Physics, McGill University, 3600 University Street, Montreal, Quebec, H3A 2T8, Canada}

\author{Jorge Noronha}
\affiliation{Instituto de F\'{i}sica, Universidade de S\~{a}o Paulo, C.P.
66318, 05315-970 S\~{a}o Paulo, SP, Brazil}

\author{Rone P.~G.~Andrade}
\affiliation{Instituto de F\'{i}sica, Universidade de S\~{a}o Paulo, C.P.
66318, 05315-970 S\~{a}o Paulo, SP, Brazil}

\author{Fr\'ed\'erique Grassi}
\affiliation{Instituto de F\'{i}sica, Universidade de S\~{a}o Paulo, C.P.
66318, 05315-970 S\~{a}o Paulo, SP, Brazil}

\date{\today}

\begin{abstract}
Bulk viscosity effects on the collective flow harmonics in heavy ion
collisions are investigated, on an event by event basis, using a newly
developed 2+1 Lagrangian hydrodynamic code named v-USPhydro which implements
the Smoothed Particle Hydrodynamics (SPH) algorithm for viscous
hydrodynamics. A new formula for the bulk viscous corrections present in the
distribution function at freeze-out is derived starting from the Boltzmann
equation for multi-hadron species. Bulk viscosity is shown to enhance the
collective flow Fourier coefficients from $v_2(p_T)$ to $v_5(p_T)$ when $%
p_{T}\sim 1-3$ GeV even when the bulk viscosity to entropy density ratio, $%
\zeta/s$, is significantly smaller than $1/(4\pi)$.
\end{abstract}

\pacs{25.75.-q,12.38.Mh, 24.10.Nz, 25.75.Ld}
\maketitle

\affiliation{Instituto de F\'{i}sica, Universidade de S\~{a}o Paulo, C.P.
66318, 05315-970 S\~{a}o Paulo, SP, Brazil}

\affiliation{Department of Physics, McGill University, 3600 University
Street, Montreal, Quebec, H3A 2T8, Canada}

\affiliation{Instituto de F\'{i}sica, Universidade de S\~{a}o Paulo, C.P.
66318, 05315-970 S\~{a}o Paulo, SP, Brazil}

\affiliation{Instituto de F\'{i}sica, Universidade de S\~{a}o Paulo, C.P.
66318, 05315-970 S\~{a}o Paulo, SP, Brazil}

\affiliation{Instituto de F\'{i}sica, Universidade de S\~{a}o Paulo, C.P.
66318, 05315-970 S\~{a}o Paulo, SP, Brazil}

\section{Introduction}

One of the most important discoveries that stemmed from the relativistic
heavy ion collision program was the discovery that the Quark-Gluon Plasma
(QGP) formed in these reactions behaves as a relativistic fluid for which
viscous effects appear to be very small \cite{Gyulassy:2004zy}. The large
degree of collectivity evidenced by the Fourier harmonics of the flow are
compatible \cite{Heinz:2013th} with viscous hydrodynamic calculations in
which the shear viscosity to entropy density ratio, $\eta/s$, nears the
uncertainty principle estimate $\sim 1/(4\pi)$ \cite%
{Danielewicz:1984ww,Kovtun:2004de}. Since $\eta/s$ becomes much larger than $%
1/(4\pi)$ in the high temperature perturbative regime \cite%
{Arnold:2000dr,Arnold:2003zc} and also at sufficiently low temperatures \cite%
{prakash}, it was suggested that in QCD this quantity should have a minimum $%
\sim 1/(4\pi)$ in between these different temperature regimes \cite%
{Hirano:2005wx,Csernai:2006zz}. Further support to this idea appeared in 
\cite{NoronhaHostler:2008ju} where it was shown that heavy resonances may
already considerably lower $\eta/s$ in the hadronic phase.

While relativistic hydrodynamical studies that include shear viscous
corrections are currently considered to be the state of the art in the field 
\cite{Schenke:2010rr,Schenke:2011bn} there is a priori no reason that
effects from bulk viscosity $\zeta $ should not also be included in the
description of the time evolution of the QGP. High temperature perturbative
QCD calculations \cite{Arnold:2006fz} show that $\zeta /s$ vanishes as $\sim
\alpha _{s}^{2}/\ln (1/\alpha _{s})$ for massless quarks and, thus, it
becomes negligible in comparison to $\eta /s$ in this regime (recent
calculations of $\zeta /s$ in a pion gas performed in the regime $T\ll m_{pion}$ can be found in \cite{Lu:2011df,Dobado:2011qu}). Given the multitude of
mass states present in QCD at intermediate temperatures in the hadronic
phase and the maximal violation of conformal invariance observed in lattice
simulations when $T\sim 150-250$ MeV \cite{Borsanyi:2010cj}, it has been
suggested that $\zeta /s$ may display a peak in the same temperature region 
\cite{Kharzeev:2007wb,Karsch:2007jc}. It is not clear at the moment if such
a peak exists in QCD, though model calculations have given support to this
idea \cite{Ozvenchuk:2012kh,Gubser:2008sz} and some of its phenomenological
consequences in heavy ion collisions have already been investigated \cite%
{Torrieri:2007fb,Torrieri:2008ip,Basar:2012bp}. The effects of bulk
viscosity in hydrodynamic simulations of the QGP have not been investigated
as thoroughly as in the case of shear viscosity. Hydrodynamical calculations
that have used nonzero $\zeta /s$ within averaged initial conditions include 
\cite%
{Monnai:2009ad,Song:2009rh,Bozek:2009dw,Denicol:2009am,Roy:2011pk,Dusling:2011fd}
while event-by-event simulations have been studied in \cite{Bozek:2012fw}.

In this paper we study the effects of bulk viscosity on the fluid-dynamical
evolution of the QGP and also at freeze-out on an event-by-event basis. The
effects of bulk viscous pressure on differential collective flow Fourier
coefficients $v_{2}$--$v_{5}$ are computed. Furthermore, a new formula for
the bulk viscous corrections in the distribution function at freeze-out is
derived starting from the Boltzmann equation for multi-particle hadron
species. Our calculations show that differential collective flow Fourier
coefficients from $v_{2}$ to $v_{5}$ are enhanced by bulk viscosity, in the
range of $p_{T}\sim 1-3$ GeV even when $\zeta /s$ is significantly smaller
than $1/(4\pi )$. Previously, this was only known to occur to $v_{2}$ (see,
for instance, \cite{Monnai:2009ad,Dusling:2011fd}). The interplay between
shear and bulk viscosities in the evolution and freeze-out of the QGP will
be investigated in a forthcoming paper.

The calculations presented in this work were performed using a new
relativistic hydrodynamics numerical code called viscous Ultrarelativistic
Smoothed Particle hydrodynamics (v-USPhydro). In this code the equations of
2+1 (i.e, boost invariant) relativistic viscous hydrodynamics are solved
using the Smoothed Particle Hydrodynamics (SPH) Lagrangian algorithm
originally developed in \cite{originalSPH,SPHothers} and later adapted in 
\cite{Aguiar:2000hw} for applications in heavy ion collisions. SPH is a mesh
free algorithm widely used in several different applications of fluid
dynamics that range from cosmology \cite{Springel:2005mi} to engineering 
\cite{sphbook}. Standard grid-based, Eulerian algorithms are known to become
very time consuming and eventually run into problems when dealing with fluid
dynamical problems involving free surfaces, deformable boundaries, and
extremely large deformations \cite{sphbook}. Given the extremely rapid time
evolution and the large gradients that appear in event by event simulations
of QGP, it is conceivable that mesh free methods such as SPH can be
instrumental in solving the equations of relativistic hydrodynamics in heavy
ion collision applications. In fact, this algorithm has been the basis for
the hydrodynamic part of the well-known NeXSPheRIO code \cite%
{Aguiar:2000hw,Osada:2001hw,Aguiar:2001ac,Socolowski:2004hw,Hama:2004rr,Hama:2005dz,Andrade:2006yh,Andrade:2008xh,Takahashi:2009na,DerradideSouza:2011rp,Gardim:2011xv,Gardim:2012yp,Gardim:2012im}%
. However, an important distinction between v-USPhydro and NeXSPheRIO,
besides the assumption of boost invariant dynamics made in v-USPhydro, is
that viscous effects in the hydrodynamical evolution and freeze-out are
included in v-USPhydro. The details of the code and the tests made to
confirm its accuracy are presented in several Appendices.

\textit{Definitions}: Our metric signature in Minkowski space-time is mostly
minus, i.e., $+,-,-,-$ and we use natural units $\hbar=k_B=c=1$. We employ
Greek indices for the 4-vectors, e.g, the 4-momentum is $p^{\mu }$,
space-time coordinates are $x^{\mu }$, the flow field $u^{\mu }$, while we
use bold letters for vectors in the transverse plane, e.g., $\mathbf{a}$,
and latin indices for the components $a_{i}$. The scalar product among
4-vectors is denoted as $p\cdot u=p_{\mu }u^{\mu }$ while for spatial
vectors we have $\mathbf{a}\cdot \mathbf{b}=a_{x}b_{x}+a_{y}b_{y}$.
Throughout this paper we will be using hyperbolic coordinates $x^{\mu
}=(\tau ,\mathbf{r},\eta )$ defined by 
\begin{eqnarray}
\tau &=&\sqrt{t^{2}-z^{2}}  \nonumber \\
\eta &=&\frac{1}{2}\ln \left( \frac{t+z}{t-z}\right) \,.
\end{eqnarray}%
The metric in hyperbolic coordinates is $g_{\mu \nu }=(1,-1,-1,-\tau ^{2})$
while the boost invariant configuration for the flow is $u_{\mu }=\left( 
\sqrt{1+u_{x}^{2}+u_{y}^{2}},u_{x},u_{y},0\right) $ and, thus, $u\cdot u=1$.

\section{The Equations of Motion of 2+1 Viscous Relativistic Hydrodynamics
Including Bulk Viscosity}

\label{eom}

We assume a vanishing baryon chemical potential and thus investigate only
the equations of motion that stem from energy and momentum conservation
within a boost invariant setup. The conservation of energy and momentum is
given by 
\begin{equation}
\frac{1}{\sqrt{-g}}\partial _{\mu }\left( \sqrt{-g}T^{\mu \nu }\right)
+\Gamma _{\lambda \mu }^{\nu }T^{\lambda \mu }=0  \label{eqn:hydro}
\end{equation}%
where $\sqrt{-g}=\tau $ and the Christoffel symbol is 
\begin{equation}
\Gamma _{\lambda \mu }^{\nu }=\frac{1}{2}g^{\nu \sigma }\left( \partial
_{\mu }g_{\sigma \lambda }+\partial _{\lambda }g_{\sigma \mu }-\partial
_{\sigma }g_{\mu \lambda }\right) .
\end{equation}%
The most general expression for the energy-momentum tensor (in the absence
of shear viscosity effects) is 
\begin{equation}
T^{\mu \nu }=\varepsilon u^{\mu }u^{\nu }-\left( p+\Pi \right) \Delta ^{\mu
\nu },
\end{equation}%
where $\Pi $ is the bulk viscous pressure and the spatial projector is $%
\Delta _{\mu \nu }=g_{\mu \nu }-u_{\mu }u_{\nu }$. Above, we use the Landau
definition for the local rest frame, $u_{\nu }T^{\mu \nu }=\varepsilon
u^{\mu }$. The remaining dynamical quantities are the energy density $%
\varepsilon $, the pressure $p$, and the fluid 4-velocity $u^{\mu }$.
Besides energy-momentum conservation, one also needs to specify the
differential equation obeyed by $\Pi $. In this paper we employ the simplest
second order formulation of the fluid dynamical equations of motion that can
be causal and stable 
\begin{equation}
\tau _{\Pi }\left( D\Pi +\Pi \theta \right) +\Pi +\zeta \theta =0,
\label{eqn:hydro3}
\end{equation}%
where $D=u^{\mu }\partial _{\mu }$ is the comoving covariant derivative, $%
\theta =\tau ^{-1}\partial _{\mu }\left( \tau u^{\mu }\right) $ is the fluid
expansion rate, $\zeta $ is the bulk viscosity, and $\tau _{\Pi }$ is the
relaxation time coefficient required to preserve causality (see, e.g., the
discussion in \cite{Denicol:2008ha}). Equation\ (\ref{eqn:hydro3}) is
discussed in detail in \cite{Denicol:2008ua,Denicol:2009am}. Note that the
differential equation for $\Pi $ in (\ref{eqn:hydro3}) includes the
nonlinear term $\Pi \theta $. For a full derivation of the equations of
motion using the method of moments for the Boltzmann equation including the
numerous other terms that enter at that order the reader is referred to \cite%
{Denicol:2012cn}. In this paper viscous effects associated with bulk
viscosity are encoded in only 2 transport coefficients, but the inclusion of
the remaining terms is straightforward and is left for a future work.

Fluid-dynamical evolution can be described using either the Eulerian or the
Lagrangian approach. Eulerian methods require the presence of a grid where
the hydrodynamical fields are defined while in Lagrangian methods the flow
is described following the trajectory of fluid \textquotedblleft particles"
and one can consistently rephrase the field equations in this language. In
cosmology applications, Lagrangian methods have become largely employed
because they allow for quick computational times and avoid other limitations
that are inherent to grid-based methods. Currently, in heavy ion collisions
almost all hydrodynamical codes are written in the Eulerian formulation with
the exception of the Lagrangian-based codes in \cite%
{Aguiar:2000hw,Nonaka:2006yn,Denicol:2009am}. The SPH formulation of the
equations of motion (\ref{eqn:hydro}-\ref{eqn:hydro3}) is reviewed in
Appendix \ref{appendixSPH}. Since the results of this paper are based on a
new relativistic fluid dynamics simulation, we describe tests made to the
algorithm in Appendix \ref{appendixSPH} and \ref{sec:convh}.

In the following, we shall always assume the Bjorken scaling solution for
the longitudinal direction, in which the component of the velocity field in
the longitudinal direction (in hyperbolic coordinates) is set to zero, i.e., 
$u^{\eta }=0\,$.

\section{Parameters of the Model}

As in any hydrodynamical modeling of heavy ion collisions, there are a
number of free parameters that must be fixed. The initial time to start the
fluid-dynamical evolution is fixed to be $\tau _{0}=1$ fm. The initial
conditions for the hydrodynamic simulations are taken from a Monte Carlo
Glauber code \cite{ic} in which the initial energy density is given in the
form 
\begin{equation}
\varepsilon (\mathbf{r})=c\;n_{coll}(\mathbf{r})
\end{equation}%
where $n_{coll}$ is the number density of binary collisions in the event and
the constant $c$ is fixed to describe the final multiplicities observed
experimentally. The centrality classes we used are defined in terms of the
number of participant nucleons and are in agreement with standard results
from other Monte Carlo Glauber simulations \cite{Adler:2003cb}. We assumed
that for $\sqrt{s}=200$ GeV RHIC's most central collisions at mid-rapidity
there are about 300 $\pi ^{+}$'s \cite{Arsene:2004fa} and, since we assume
that freeze-out occurs at $T=150$ MeV, we estimate that about $41\% $ of
pions would come from direct thermal pions \cite{jakiphd}. Thus, the
constant $c$ was adjusted for the ideal case using an average 0-5$\%$ most
central RHIC event so that we obtain $123$ $\pi ^{+}$'s when $T_{FO}=150$
MeV. Note that when bulk viscosity is included the total entropy of the
system increases in time and, in order to keep the total number of pions at
freeze-out the same as before, one has to slightly reduce $c$. Here, we
assume that $\Pi $ and the spatial components of $u^{\mu }$ vanish at $\tau
_{0}$.\ 

Initial conditions that include the effects of gluon saturation, such as
those in \cite{ic,Gale:2012rq,Moreland:2012qw}, can also be studied using
v-USPhydro. However, this type of initial conditions displays structure on
smaller length scales than that of the usual Glauber initial conditions and,
thus, smaller values of the smoothing parameter $h$ (more on that below)
must be used to systematically investigate the difference created in the
flow harmonics due to gluon saturation effects at early times.

We use the lattice-based equation of state from \cite{Huovinen:2009yb} with
chemical equilibrium. We have tested the dependence of our results with the
choice for the equation of state by using the equation of state from Fodor
et al. \cite{Borsanyi:2010cj} which, however, led to no noticeable
difference in the flow harmonics. The results shown in this paper were
computed using EOS S95n-v1 \cite{Huovinen:2009yb} (for $T < 50$ MeV this EOS
is matched to that of a massive gas of pions).

The bulk viscosity coefficient is parametrized in the following way 
\begin{equation}
\frac{\zeta }{s}=\frac{1}{8\pi }\left( \frac{1}{3}-c_{s}^{2}\right) ,
\label{eqn:adszeta}
\end{equation}%
where $s$ is the entropy density and $c_{s}$ is the speed of sound. As one
can see in Fig.\ 1, $\zeta /s$ is significantly smaller than the standard
value of $\eta /s\sim 1/(4\pi )$. Additionally, because of the dip in the
speed of sound, there is a peak in $\zeta /s$ between $150-200$ MeV. This
formula for the bulk viscosity is inspired by Buchel's formula obtained
within the gauge/gravity duality \cite{Buchel:2007mf}. Calculations
performed in \cite{Arnold:2006fz} showed that at sufficiently large $T$ in
the QGP phase the bulk viscosity in QCD follows the relation $\zeta /\eta
\sim 15\left(1/3-c_{s}^{2}\right)^{2}$. The specific functional form of $%
\zeta /s$ does not make a significant difference in our results as long as
the overall magnitude is the same.


The temperature dependence of the relaxation time, $\tau _{\Pi }$, is
described by the formula found in \cite{Huang:2010sa} 
\begin{equation}
\tau _{\Pi }=9\,\frac{\zeta }{\varepsilon -3p}\,.  \label{eqn:taupi}
\end{equation}%
and shown in Fig.\ 1. Clearly, other choices for $\tau _{\Pi }$ are possible
but, for simplicity, in this paper we fix it as above. We chose a time step $%
d\tau =0.1$ fm in our numerical simulations and $\tau _{\Pi }$ is set to
never be smaller than $d\tau$ (this is why $\tau_{\Pi}$ plateaus when $T >
0.2$ GeV). Also, the time step is considerably smaller than $\tau _{0}$, as
required to resolve the gradients in the longitudinal direction.

\begin{figure}[tbp]
\begin{tabular}{cc}
\includegraphics[width=0.4\textwidth]{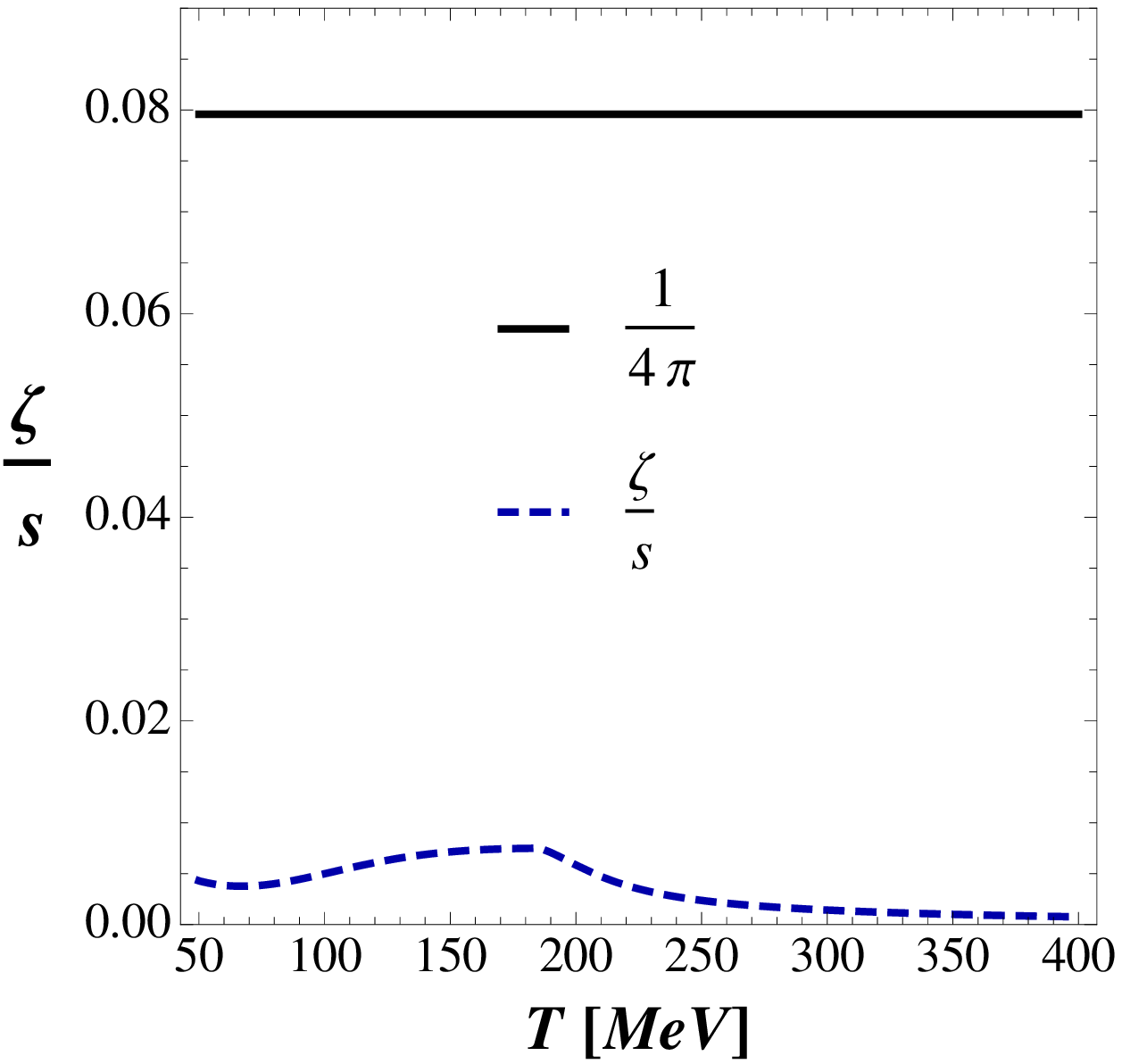} & %
\includegraphics[width=0.4\textwidth]{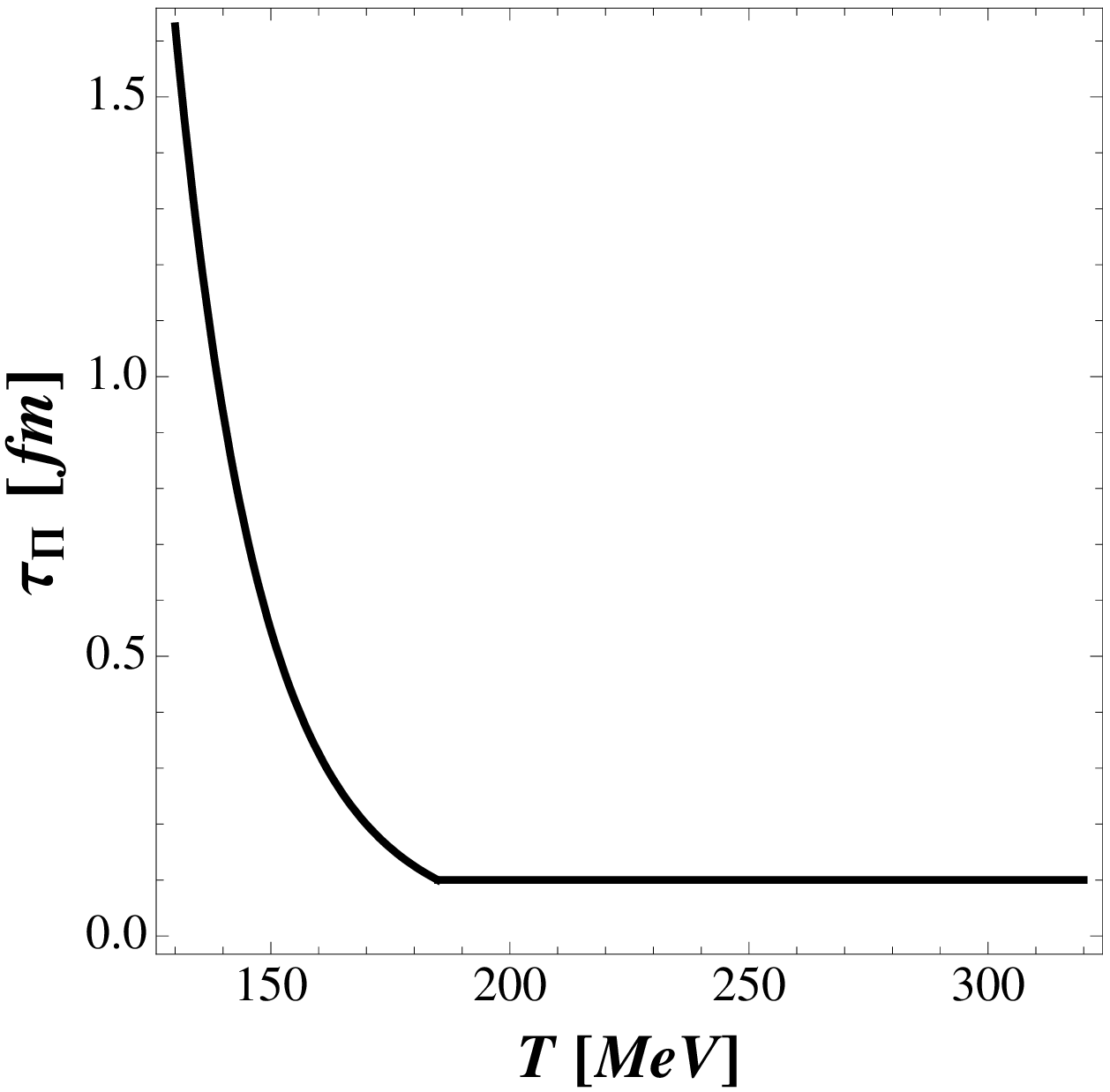} \\ 
& 
\end{tabular}%
\caption{The temperature dependence of the transport coefficients $\protect%
\zeta/s$ and $\protect\tau_{\Pi}$ (see Eqs.\ (\protect\ref{eqn:adszeta}) and
(\protect\ref{eqn:taupi}), respectively).}
\label{tab:vint}
\end{figure}

We set the isothermal freeze-out temperature in the Cooper-Frye procedure 
\cite{Cooper:1974mv} to be $T_{FO}=150$ MeV. The Cooper-Frye freeze-out
method is written in the SPH language in Appendix \ref{cooperfryeappendix}.
Lower freeze-out temperatures have a strong effect on the viscous
corrections to the Cooper-Frye freeze-out, as it will be explained in the
next section. On the other hand, higher freeze-out temperatures can run into
issues because of the effects coming from the production of heavy resonances
that could contribute strongly to the flow harmonics \cite%
{Noronha-Hostler:2013ria}. In this paper all flow harmonics are shown for
direct thermal $\pi^{+}$, without the inclusion of a hadronic afterburner or
particle decays. Thus, the purpose of the present study is to understand the
qualitative effects of bulk viscosity on the flow harmonics event by event.
In the future, we will include particle decays (and possibly hadronic
afterburner effects) into v-USPhdro so that the results of our calculations
can be more directly compared to experimental results.

As explained in detail in Appendix \ref{appendixSPH}, the Lagrangian
formalism we used includes a parameter, $h$, known as the SPH length scale.
The $h$ parameter essentially works as a smoothing parameter such that a
larger $h$ smoothes out initial conditions to the point that fluctuations are
minimized whereas a too small $h$ requires a very large number of SPH
particles so that it overwhelms standard computational times. Thus, one
needs to choose a value of $h$ that is small enough to preserve as much of
the structure in the initial conditions as possible but that also allows for
a realistic running time. We found that $h=0.3$ fm and $N_{SPH}=27000$ allow
for relatively quick running times ($\sim $ 10 minutes per hydro event in a
single standard machine) while still preserving the structure seen in the
Monte Carlo Glauber initial fluctuations for both centrality classes studied
in this paper. If one increases $h$ beyond that value the initial
fluctuations are smoothed out and we see $v_{2}$ increase by about $5\%$ at
high $p_{T}$ when one uses $h=0.5$ fm. For further details and discussion
see Appendix \ref{sec:convh}. Conservation of energy and momentum in our
simulations hold up to $0.01\%$.

\section{Bulk viscosity contribution to the multi-hadron distribution
function}

\subsection{Method of Moments}

In order to compute the particle distribution at freeze-out, one needs to
evaluate the non-equilibrium distribution function for each hadron species
on the freeze-out hypersurface. In general, the distribution function for
each hadron species is $f_{\mathbf{k}}^{(i)}=f_{0\mathbf{k}}^{(i)}+\delta f_{%
\mathbf{k}}^{(i)}$, where $f_{0\mathbf{k}}^{(i)}=[\exp (\beta _{0}E_{i\,%
\mathbf{k}}-\alpha_{0})+a]^{-1}$ is the local equilibrium distribution
function and $\delta f_{\mathbf{k}}^{(i)}$ is the corresponding
non-equilibrium part. In this work, we use the method of Moments, as
developed in Refs.~\cite{Denicol:2012cn,Denicol:2012yr}, to compute the
non-equilibrium contribution $\delta f_{\mathbf{k}}^{\left( i\right) }$
associated with bulk viscosity effects to the momentum distribution function
of a hadronic mixture.

First, we factorize $\delta f_{\mathbf{k}}^{\left( i\right) }$ in the
following way 
\[
\delta f_{\mathbf{k}}^{(i)}=f_{0\mathbf{k}}^{(i)}\tilde{f}_{0\mathbf{k}%
}^{(i)}\phi _{\mathbf{k}}^{(i)}, 
\]%
where $\tilde{f}_{0\mathbf{k}}^{(i)}=1+a f_{0\mathbf{k}}^{(i)}$ ($a=-1$/$1$
for fermions/bosons), and $\phi _{\mathbf{k}}^{(i)}$ is an
out-of-equilibrium contribution. Next, $\phi _{\mathbf{k}}^{(i)}$ is
expanded in terms of its moments using a complete and orthogonal basis
constructed from particle four-momentum, $k_{i}^{\mu }$, and fluid
four-velocity, $u^{\mu }$. As done in Refs.~\cite%
{Denicol:2012cn,Denicol:2012yr}, we use an expansion basis with two basic
ingredients: the irreducible tensors 
\[
1,k_{i}^{\left\langle \mu \right\rangle },k_{i}^{\left\langle \mu \right.
}k_{i}^{\left. \nu \right\rangle },k_{i}^{\left\langle \mu \right.
}k_{i}^{\nu }k_{i}^{\left. \lambda \right\rangle },\cdots \ , 
\]%
analogous to the well-known set of spherical harmonics and constructed by
the symmetrized traceless projection of $k_{i}^{\mu _{1}}\cdots k_{i}^{\mu
_{m}}$, i.e., $k_{i}^{\left\langle \mu _{1}\right. }\cdots k_{i}^{\left. \mu
_{m}\right\rangle }\equiv \Delta _{\nu _{1}\cdots \nu _{m}}^{\mu _{1}\cdots
\mu _{m}}k_{i}^{\nu _{1}}\cdots k_{i}^{\nu _{m}}$, and the orthonormal
polynomials 
\[
P_{i\mathbf{k}}^{\left( n\ell \right) }=\sum_{r=0}^{n}a_{nr}^{(\ell
)i}\left( u\cdot k_{i}\right) ^{r}, 
\]%
which are equivalent to the associated Laguerre polynomials in the limit of
massless, classical particles.

Then, the momentum distribution function of the $i$--th particle species
becomes,

\begin{equation}
f_{\mathbf{k}}^{(i)}=f_{0\mathbf{k}}^{(i)}+f_{0\mathbf{k}}^{(i)}\tilde{f}_{0%
\mathbf{k}}^{(i)}\sum_{\ell =0}^{\infty }\sum_{n=0}^{\infty }\mathcal{H}_{i%
\mathbf{k}}^{\left( n\ell \right) }\rho _{i,n}^{\mu _{1}\cdots \mu _{\ell
}}k_{i,\mu _{1}}\cdots k_{i,\mu _{\ell }},  \label{expansion}
\end{equation}%
where we introduced the energy-dependent coefficients, $\mathcal{H}_{i%
\mathbf{p}}^{\left( n\ell \right) }\equiv \left[ N_{i}^{\left( \ell \right)
}/\ell !\right] \sum_{m=n}^{\infty }a_{mn}^{(\ell )i}P_{i\mathbf{k}}^{\left(
m\ell \right) }\left( u\cdot k_{i}\right) $ (see the details in \cite%
{Denicol:2012cn}). The fields $\rho _{i,n}^{\mu _{1}\cdots \mu _{\ell }}$
can be exactly determined using the orthogonality relations satisfied by the
expansion basis and can be shown to correspond to irreducible moments of $%
\delta f_{\mathbf{k}}^{(i)}$,%
\begin{equation}
\rho _{i,r}^{\mu _{1}\ldots \mu _{\ell }}\equiv \left\langle E_{i\mathbf{k}%
}^{r}k_{i}^{\left\langle \mu _{1}\right. }\ldots k_{i}^{\left. \mu _{\ell
}\right\rangle }\right\rangle _{\delta }\text{, \ \ \ }\left\langle \ldots
\right\rangle _{\delta }=\int dK_{i}\text{ }\left( \ldots \right) \delta f_{%
\mathbf{k}}^{(i)},  \label{Hk}
\end{equation}%
where $g_{i}$ is the degeneracy factor of the $i$--th hadron species and $%
dK_{i}=g_{i}d^{3}\mathbf{k/}\left[ \left( 2\pi \right) ^{3}k_{i}^{0}\right] $%
. As long as this basis is complete, the above expansion fully describes $f_{%
\mathbf{k}}^{(i)}$, no matter how far from equilibrium the system is.

Here, we are interested only on the effects arising from the bulk viscous
pressure. For this case, it is enough to consider only the $\ell =0$
(scalars) terms in the expansion above, i.e., neglect all irreducible
first-rank tensors, e.g. heat flow, second-rank tensors, e.g. shear-stress
tensor, and tensors with rank higher than two (that never appear in fluid
dynamics). The next approximation is the truncation of the expansion in
momentum space, keeping only the terms corresponding to $n=0,1,2$ (for $\ell
=0$ ). Then, we obtain (for classical particles)%
\begin{eqnarray}
f_{\mathbf{k}}^{(i)} &=&f_{0\mathbf{k}}^{(i)}+\delta f_{\mathbf{k}}^{(i)}%
\text{ }, \\
\delta f_{\mathbf{k}}^{(i)} &=&\frac{f_{0\mathbf{k}}^{i}}{J_{00}^{i}}\left\{ %
\left[ 1+a_{10}^{(0)i}a_{10}^{(0)i}+a_{20}^{(0)i}a_{20}^{(0)i}+\left(
a_{10}^{(0)i}a_{11}^{(0)i}+a_{20}^{(0)i}a_{21}^{(0)i}\right) u\cdot
k_{i}+a_{20}^{(0)i}a_{22}^{(0)i}\left( u\cdot k_{i}\right) ^{2}\right] \rho
_{i\left( 0\right) }\right. \; \\
&&\left. +a_{22}^{(0)i}\text{ }\left[ a_{20}^{(0)i}+a_{21}^{(0)i}u \cdot
k_{i}+a_{22}^{(0)i}\left( u \cdot k_{i}\right) ^{2}\right] \rho _{i\left(
2\right) }\right\} .
\end{eqnarray}

The coefficients $J_{mn}^{i}$ and $a_{mn}^{(0)i}$ are thermodynamic
quantities that appear in the definition of the orthogonal polynomicals $P_{i%
\mathbf{k}}^{\left( n\ell \right) }$. These functions are defined in
Appendix \ref{Deltaf_Coeffs}. Note that $m_{i}^{2}\rho _{i,0}=\rho
_{i,2}-3\Pi _{i}$, with $\Pi _{i}$ being the bulk viscous pressure \textit{%
of the }$i$\textit{--th particle species}, respectively. Likewise, $%
\varepsilon _{i}$, and $p_{i}$ are the energy density and thermodynamic
pressure of the $i$--th particle species, respectively.

In order to apply this expression to describe freeze-out, further
approximations are required. This happens because in fluid dynamics one
evolves the total bulk viscous pressure of the system ($\Pi =\sum_{i}\Pi
_{i} $ ) and it is not possible to know, just from fluid dynamics itself,
how these quantities are distributed among the individual bulk viscous
pressure of each hadron species ($\Pi _{i}$). We remark that $\Pi
=-\sum_{i}\left( m_{i}^{2}/3\right) \rho _{i,0}$ and $\sum_{i}\rho _{i,2}=0$%
, the second relation arising from the Landau matching condition.\
Therefore, we need to relate the scalar moments $\rho _{i,0}$, $\rho _{i,1}$%
, and $\rho _{i,2}$ to the fluid-dynamical variables $\Pi $.

In this work, we assume that the system is close to the Navier-Stokes limit.
Even though this assumption does not happen in practice, we consider it good
enough to provide a rough estimate for the non-equilibrium distribution
function. In this case,%
\[
\Pi =-\zeta \theta ,\text{ }\rho _{i,m}=-\alpha _{i,m}\theta \Longrightarrow
\rho _{i,m}=\frac{\alpha _{i,m}}{\zeta }\Pi . 
\]

Then%
\begin{eqnarray}
\delta f_{\mathbf{k}}^{(i)} &=&\frac{f_{0\mathbf{k}}^{i}}{J_{00}^{i}}\Pi
\left\{ \left[ 1+a_{10}^{(0)i}a_{10}^{(0)i}+a_{20}^{(0)i}a_{20}^{(0)i}+%
\left( a_{10}^{(0)i}a_{11}^{(0)i}+a_{20}^{(0)i}a_{21}^{(0)i}\right) u\cdot
k_{i}+a_{20}^{(0)i}a_{22}^{(0)i}\left( u \cdot k_{i}\right) ^{2}\right] 
\frac{\alpha _{i,0}}{\zeta }\right. \; \\
&&\left. +a_{22}^{(0)i}\text{ }\left[ a_{20}^{(0)i}+a_{21}^{(0)i}u \cdot
k_{i}+a_{22}^{(0)i}\left( u \cdot k_{i}\right) ^{2}\right] \frac{\alpha
_{i,2}}{\zeta }\right\} , \\
&=&f_{0\mathbf{k}}^{i}\Pi \left[ B_{0}^{(i)}+D_{0}^{(i)}u \cdot
k_{i}+E_{0}^{(i)}\left( u \cdot k_{i}\right) ^{2}\right] .
\end{eqnarray}%
The coefficients $B_{0}^{(i)}$, $D_{0}^{(i)}$, and $E_{0}^{(i)}$ are defined
as implied. In general, while these coefficients depend on the freeze-out
temperature and the particle's mass and degeneracy, they cannot be simply
expressed in terms of thermodynamical quantities such as energy density per
particle and etc.

Note that the dependence of $\delta f^{(i)}$ with the particle's 4-momentum
is qualitatively different than that commonly used for shear viscous effects
where $\delta f_{shear} \sim \pi_{\mu\nu}k^\mu_i
k^\nu_i/(\varepsilon_i+p_i)/T^2$, where $\pi^{\mu\nu}$ is the shear tensor 
\cite{Denicol:2012yr}. Taking the equivalent Navier-Stokes limit for the
shear correction, one can see the relevant quantity in this case is $\eta/s$%
. Thus, if $\eta/s$ is sufficiently small, the $\delta f$ expansion may
still be well defined for intermediate values of $p_T$. However, for bulk
viscosity, even in the Navier-Stokes limit the relevant quantity in the $%
\delta f$ series is not $\zeta/s$. Rather, one can see that at low momenta
the relevant quantity is $\sim \zeta \theta B_0$ while as one increases the
momentum the important quantity becomes $\sim E_0 \left( u \cdot
k_{i}\right) ^{2} \zeta \theta$. Therefore, a relatively small $\zeta/s$ (in
comparison to $\eta/s$, for example) does not necessarily mean that the
actual values that matter in the viscous contribution to freeze-out are
going to be small.

\subsection{Simple Model of Hadrons}

In order to compute the coefficients $\alpha _{i,m}$ appearing in $\delta f_{%
\mathbf{k}}^{(i)}$ we must provide a set of hadronic cross sections. In this
work, we estimate these coefficients using a simple hadronic model in which
all hadrons have the same constant cross section. Note that in this case the
ratios $\alpha _{i,0}/\zeta $ and $\alpha _{i,2}/\zeta $ are actually
independent of the value of cross section chosen. We consider only elastic
collisions between the hadrons and include all hadrons up to a mass of $1.2$
GeV (heavier hadrons are not included due to the exceedingly large
computational cost).

Since we consider only elastic collisions among the hadrons, the particle
number of individual species is conserved and the moments $\rho _{i,1}$
vanish, $\rho _{i,1}=0$. In this case, the coefficients $B_{0}$, $D_{0}$,
and $E_{0}$ for pions with freeze-out temperature $T_{FO}=150$ MeV are 
\begin{eqnarray}
B_{0}^{(\pi)}&=& -65.85 \,\,fm^4\,,  \nonumber \\
D_{0}^{(\pi)}&=& 171.27 \,\,fm^4/GeV\,,  \nonumber \\
E_{0}^{(\pi)}&=& -63.05 \,\,fm^4/GeV^2\,,
\end{eqnarray}
and the non-equilibrium correction in the distribution function for pions is 
\begin{equation}
\delta f_{\mathbf{k}}^{(\pi)}=f_{0\mathbf{k}}^{\pi}\,\Pi \left[
B_{0}^{(\pi)}+D_{0}^{(\pi)}u\cdot k_{\pi}+E_{0}^{(\pi)}\left( u\cdot
k_{\pi}\right)^{2}\right]\,.  \label{bulkmoments}
\end{equation}
For freeze-out temperatures lower than $150$ MeV, the $\delta f$
contribution to the distribution function can become comparable to the ideal
distribution $f_0$, which makes a perturbative analysis of the viscous
effects at freeze-out untrustworthy. Thus, we used $T_{FO}=150$ MeV for the
calculations in this paper.

The Moments method \cite{Denicol:2012cn,Denicol:2012yr} can be used to
derive a relativistic dissipative fluid-dynamical theory from kinetic theory
which provides a good description of all dissipative phenomena. This has
been explicitly demonstrated in \cite{Denicol:2012vq,Greif:2013bb} where
calculations performed within this theory were shown to match the
corresponding numerical solutions of the relativistic Boltzmann equation.
However, in order to estimate the systematic uncertainties in our
calculation due to the approximations performed in this section in the
determination of $\delta f$, we also consider two other options. The first
one corresponds to the result derived by Monnai and Hirano in \cite%
{Monnai:2009ad} using an implementation of Grad's 14-moment method for
multi-particle species. The other formula for $\delta f$ was obtained by
Dusling and Sch\"afer in \cite{Dusling:2011fd} using the relaxation time
approximation and the assumption that the deviation from equilibrium for
each hadron species comes the near-zero mode similar to that found in scalar
field theory \cite{Lu:2011df}. The formulas for $\delta f$ computed in these
works can be put in the form (\ref{bulkmoments}) with the coefficients: 
\begin{eqnarray}
B_{0}^{(\pi)}&=& -0.69 \,\,fm^4\,,  \nonumber \\
D_{0}^{(\pi)}&=& -38.96 \,\,fm^4/GeV\,,  \nonumber \\
E_{0}^{(\pi)}&=& 49.69 \,\,fm^4/GeV^2\,,
\end{eqnarray}
for \cite{Monnai:2009ad} and 
\begin{eqnarray}
B_{0}^{(\pi)}&=& -71.96 \,\,fm^4\,,  \nonumber \\
D_{0}^{(\pi)}&=& 121.50 \,\,fm^4/GeV\,,  \nonumber \\
E_{0}^{(\pi)}&=& 0 \,
\end{eqnarray}
for \cite{Dusling:2011fd}.

\begin{figure}[ht]
\centering
\includegraphics[width=0.4\textwidth]{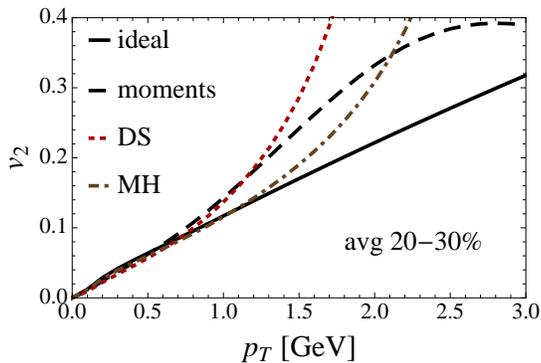}
\caption{Dependence of the direct $\protect\pi^+$ differential elliptic flow
on the specific formula for the viscous $\protect\delta f$ contribution from
the bulk viscosity that enters in the Cooper-Frye freeze-out. The results
are for RHIC's $\protect\sqrt{200}$ GeV $20-30\%$ most central collisions in
the case where the initial condition corresponds to a single average Glauber
initial condition. The ideal case is the solid black line, our result for $%
v_2$ computed using the $\protect\delta f$ obtained via the Moments method
is the long dashed black line, results for the $\protect\delta f$ described
in \protect\cite{Dusling:2011fd} is the short dashed red curve, while the
short and long dashed brown curve is the result computed using the $\protect%
\delta f$ described in \protect\cite{Monnai:2009ad}.}
\label{fig:dfcomp}
\end{figure}

In Fig.\ \ref{fig:dfcomp} the variation of $v_{2}(p_{T})$ with the different
Ans\"{a}tze for the $\delta f$ contributions from the bulk viscosity is
shown. The results are for mid-rapidity RHIC's $\sqrt{200}$ GeV $20-30\%$
most central collisions in the case where the initial condition corresponds
to a single average (over 150 events) Glauber initial condition. The ideal
case is the solid black line, our result for $v_2$ computed using the $%
\delta f$ obtained via the Moments method is the long dashed black line,
results for the $\delta f$ described in \cite{Dusling:2011fd} is the short
dashed red curve, while the short and long dashed brown curve is the result
computed using the $\delta f$ described in \cite{Monnai:2009ad}. One can see
that all the different approaches lead to an enhancement of $v_2(p_T)$ when $%
p_T \sim 1-3$ GeV in comparison to the ideal fluid case. However, note that $%
v_2(p_T)$ is more well behaved in the intermediate $1-3$ GeV range when the
result for $\delta f$ computed within the Moments method developed in \cite%
{Denicol:2012cn,Denicol:2012yr} is used in comparison to the expressions
obtained in \cite{Monnai:2009ad} and \cite{Dusling:2011fd}. In the next
section, the newly developed formula in Eq.\ (\ref{bulkmoments}) will be
used to study the effects of bulk viscosity on the collective flow harmonic
coefficients.

\section{Results for the collective flow coefficients}

\label{results}

\begin{figure}[ht]
\centering
\includegraphics[width=0.4\textwidth]{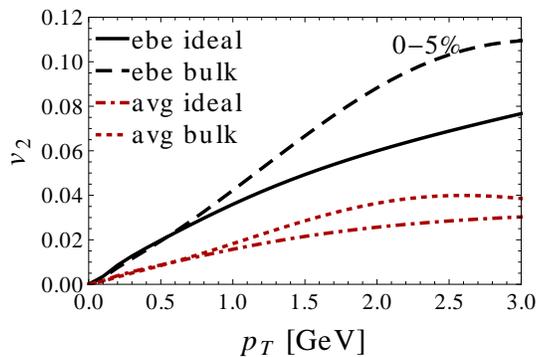}
\caption{Comparison of the effects of bulk viscosity in event-by-event
Glauber initial conditions vs. averaged Glauber initial conditions for
RHIC's $\protect\sqrt{s}=200$ GeV 0-5\% most central collisions. We have
considered 150 events. For event by event simulations, the black solid line
shows $v_2(p_T)$ for an ideal fluid while the long dashed black line shows
the effect of bulk viscosity both in the hydrodynamical evolution and
freeze-out. In the case of a single averaged Glauber simulation, the red
dashed dotted line shows $v_2(p_T)$ for an ideal fluid while the short
dashed red line shows the effect of bulk viscosity both in the
hydrodynamical evolution and freeze-out.}
\label{fig:ebe_v_avg}
\end{figure}

In this section we present our results for the effects of bulk viscosity on
the collective flow coefficients associated with direct, thermal $\pi^+$.
The flow coefficients were computed event by event using the event plane
method \cite{Poskanzer:1998yz} described for completeness in Appendix \ref%
{eventplanemethod}. In the following all results were computed for $p_{T}=0-3
$ GeV for RHIC's $\sqrt{s}=200$ GeV most central collisions ($0-5\%$) and
peripheral collisions ($20-30\%$). For each centrality class we have
considered 150 events. In Fig.\ \ref{fig:ebe_v_avg} we show a comparison of
the effects of bulk viscosity in event-by-event Glauber initial conditions
vs. averaged Glauber initial conditions for the 0-5\% most central
collisions. For event by event initial conditions, the black solid line
shows $v_2(p_T)$ for an ideal fluid while the long dashed black line shows
the effect of bulk viscosity from Eq.\ (\ref{bulkmoments}) both in the
hydrodynamical evolution and freeze-out. For a single averaged Glauber
simulation, the red dashed dotted line shows $v_2(p_T)$ in the case of an
ideal fluid while the short dashed red line shows the effect of bulk
viscosity both in the hydrodynamical evolution and freeze-out. As one can
see in Fig.\ \ref{fig:ebe_v_avg}, bulk viscosity enhances $v_2(p_T)$ when $%
p_T \sim 1-3$ GeV with respect to the ideal fluid result and this
enhancement is actually more pronounced in event by event simulations. We
shall see in the following that this enhancement is also present for higher
flow harmonics.

\begin{figure}[ht]
\centering
\begin{tabular}{cc}
\includegraphics[width=0.4\textwidth]{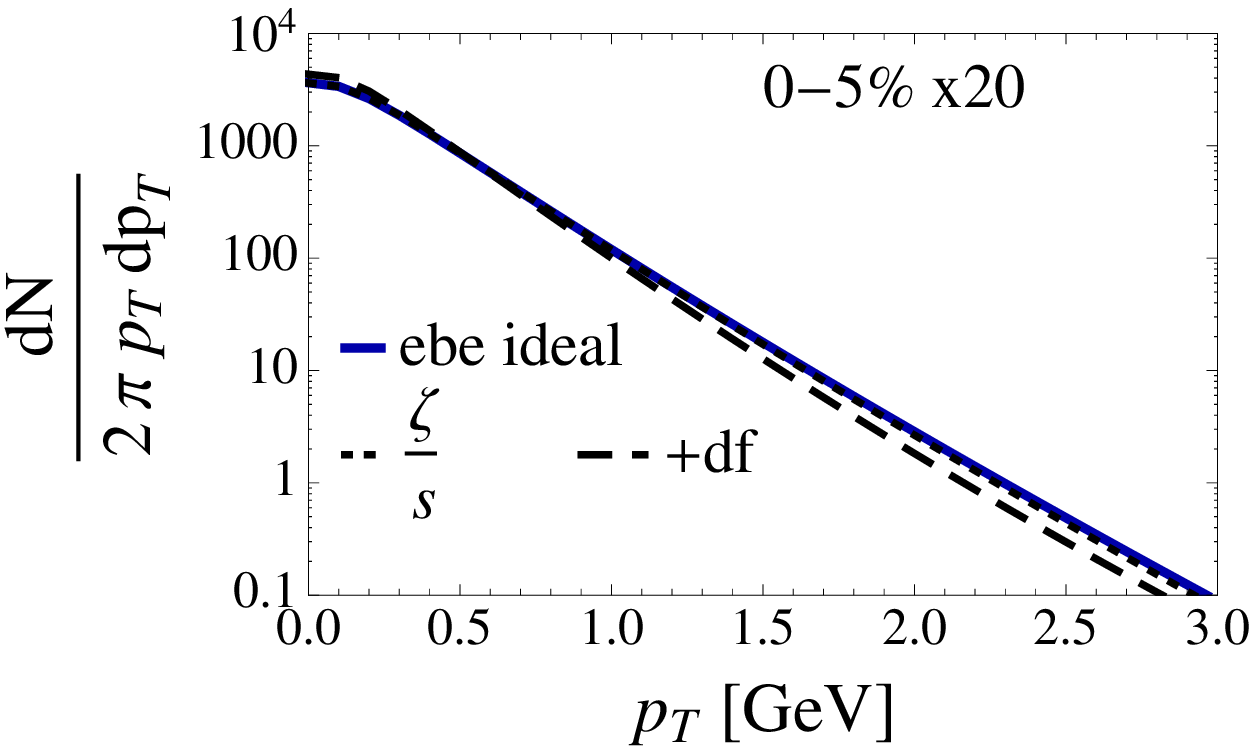} & %
\includegraphics[width=0.4\textwidth]{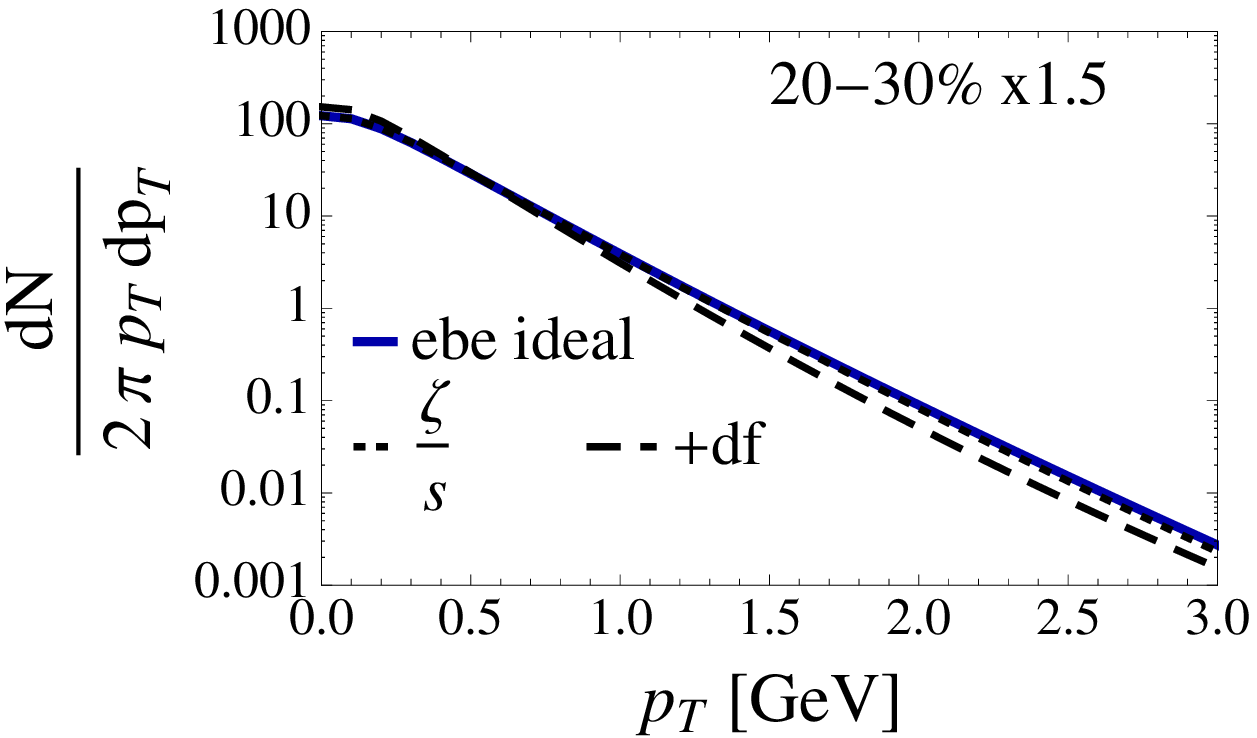} \\ 
& 
\end{tabular}%
\caption{The $\protect\pi^+$ spectra $dN/(2\protect\pi p_Tdp_T)$ for $0-5\%$
and $20-30\%$ centrality classes. The ideal fluid case is shown in a solid
blue line, the result in the case where effects of bulk viscosity are
included only on the hydrodynamical evolution but not on the freeze-out is
shown by the short dashed black line while the long dashed black curve
includes bulk effects on both the hydro evolution and freeze-out.}
\label{tab:spectra}
\end{figure}

Results for the $p_T$ spectra of $\pi^+$, $dN/(dy p_{T}dp_{T})$, are shown
in Fig.\ \ref{tab:spectra} for $0-5\%$ and $20-30\%$ centrality classes.
Effects from particle decays are not included in any of the plots shown in
this paper. One can see that bulk viscosity steepens the spectra in both
centrality classes compared to the ideal case, as can be guessed from Eq.\ (%
\ref{bulkmoments}). This is responsible for the enhancement in $v_2(p_T)$ at
high $p_T$, as remarked by \cite{Monnai:2009ad} and \cite{Dusling:2011fd}.
Note that when the $\delta f$ correction is not included in the Cooper-Frye
procedure (i.e., $\delta f=0$), the effect of bulk viscosity on spectra is
very small in both centrality classes. 

\begin{figure}[ht]
\centering
\begin{tabular}{cc}
\includegraphics[width=0.4\textwidth]{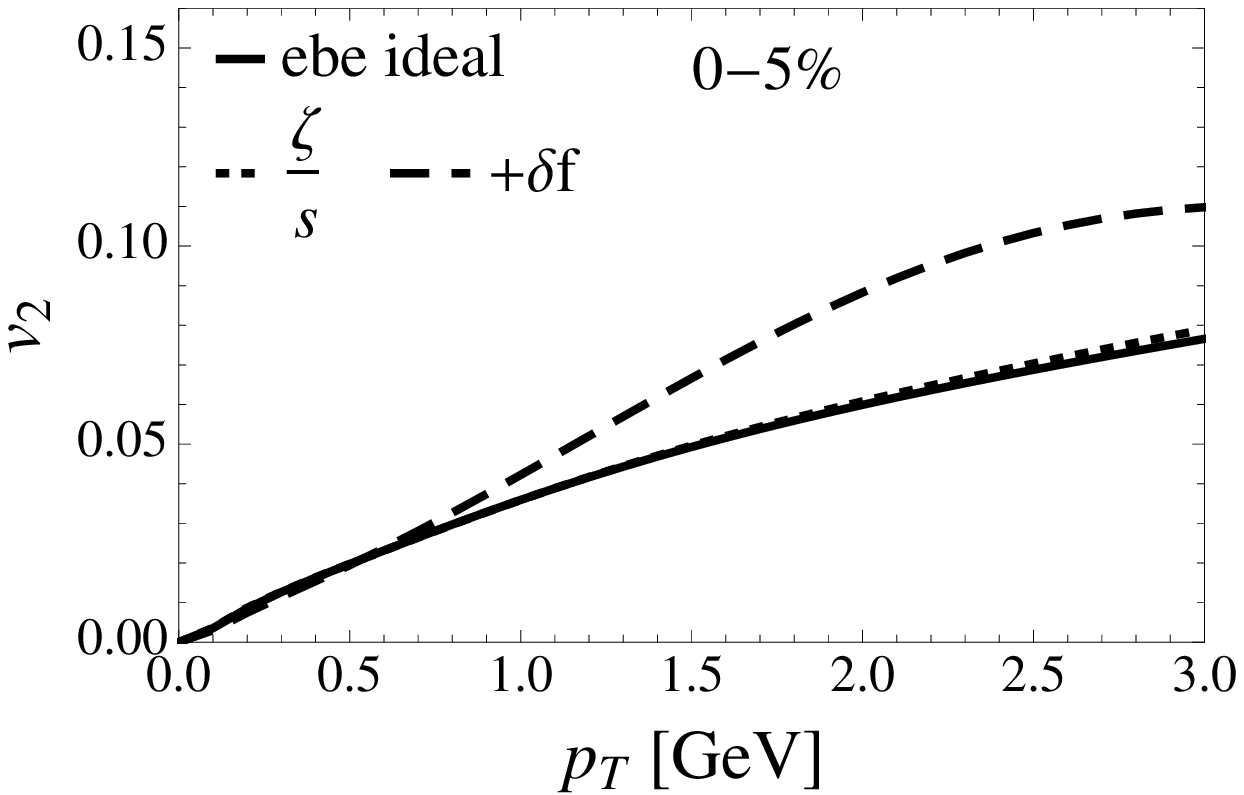} & %
\includegraphics[width=0.4\textwidth]{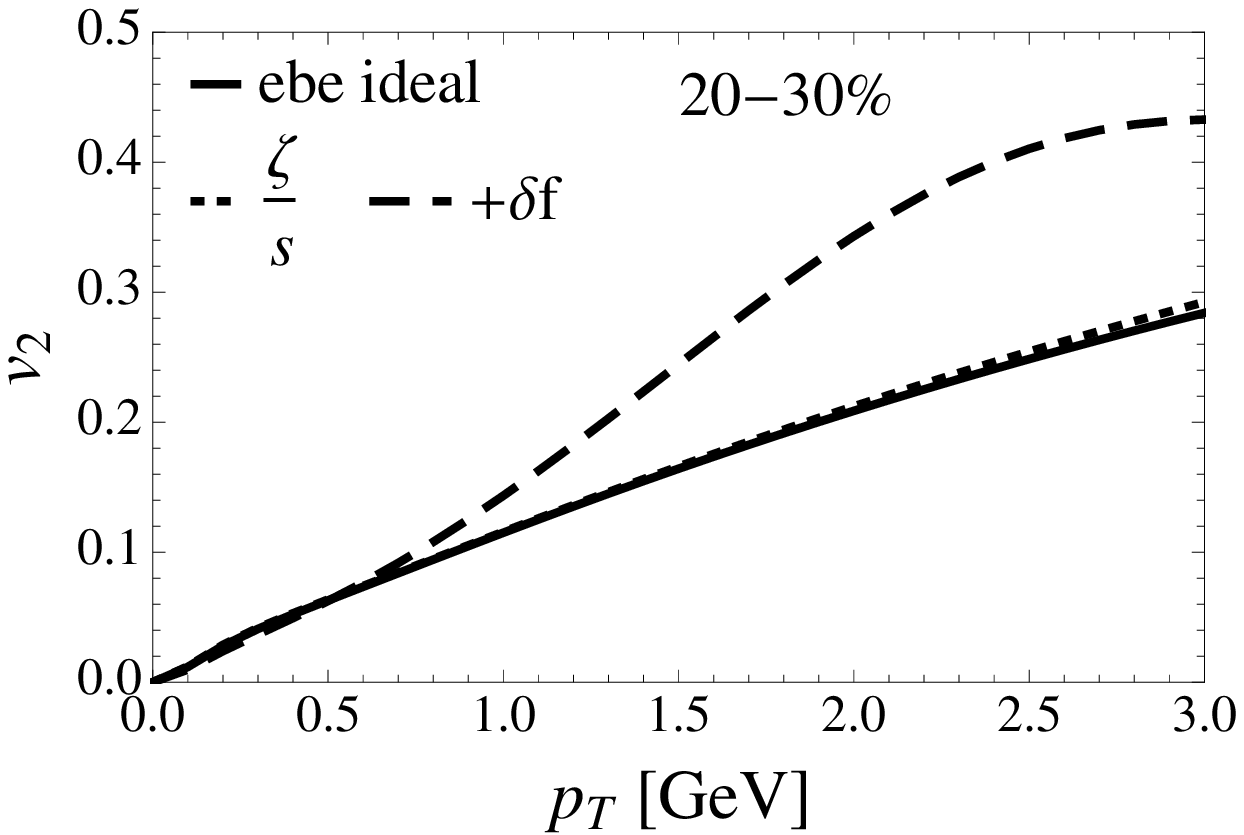} \\ 
\newline
\includegraphics[width=0.4\textwidth]{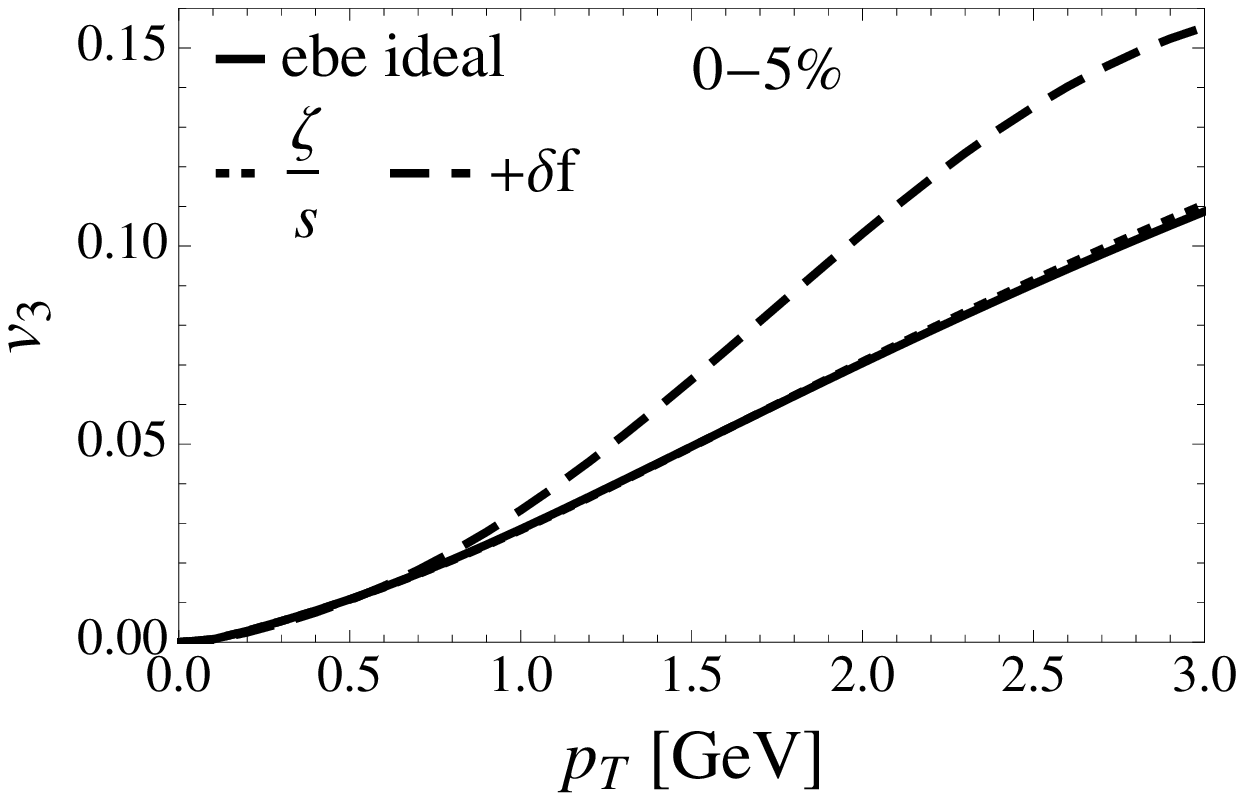} & %
\includegraphics[width=0.4\textwidth]{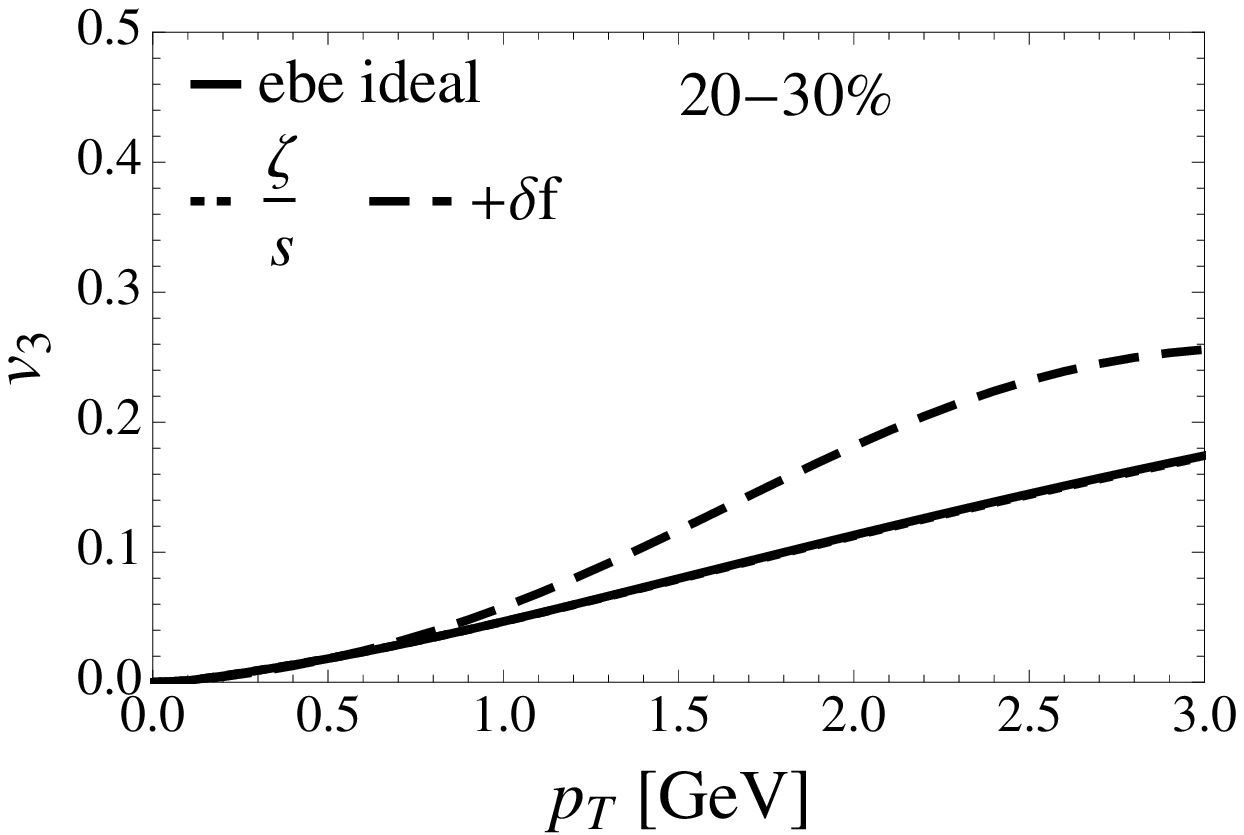} \\ 
\newline
\includegraphics[width=0.4\textwidth]{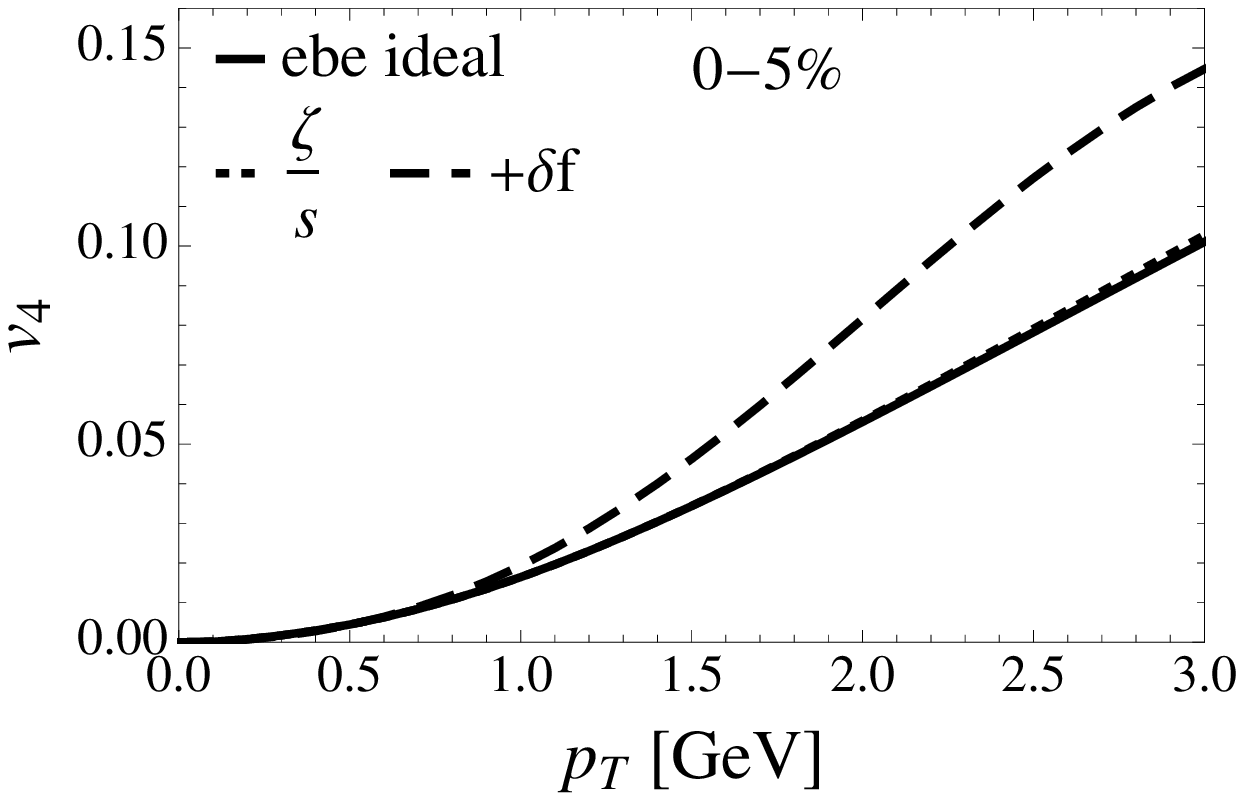} & %
\includegraphics[width=0.4\textwidth]{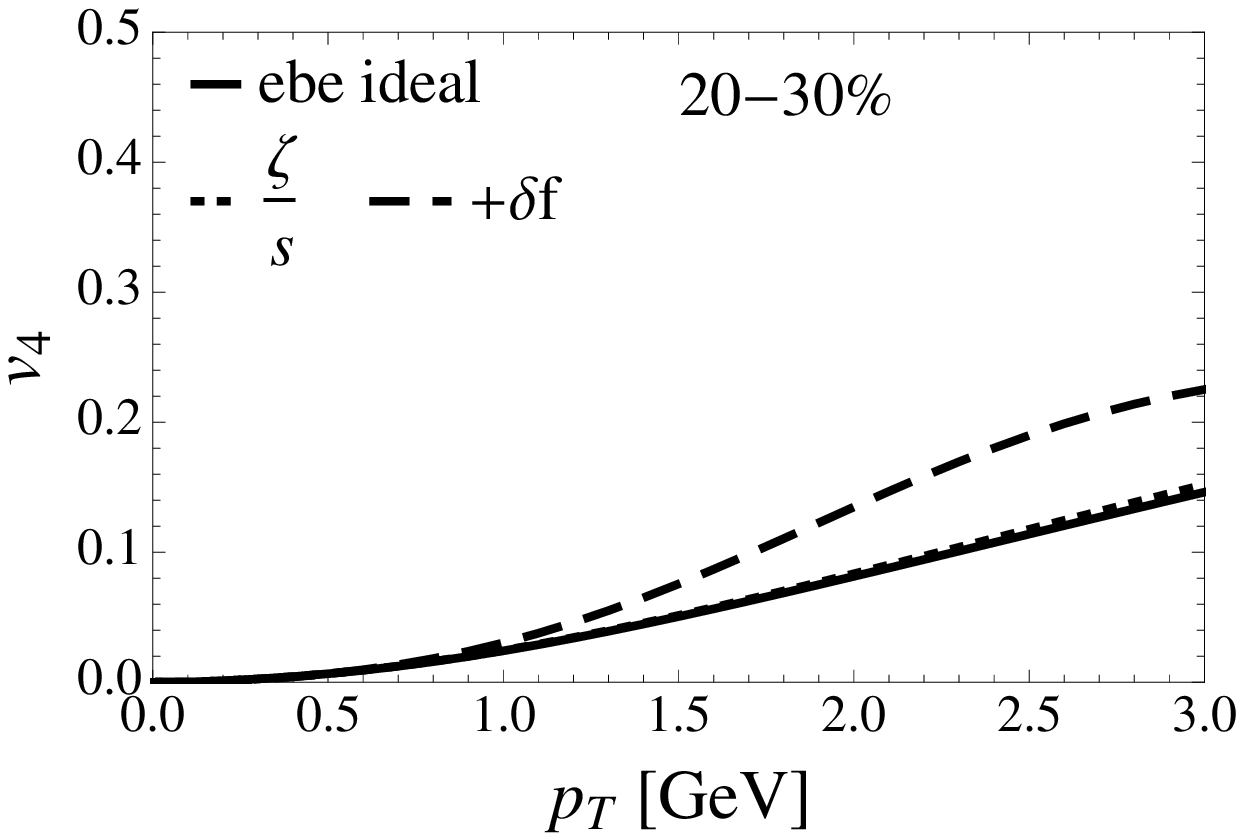} \\ 
\newline
\includegraphics[width=0.4\textwidth]{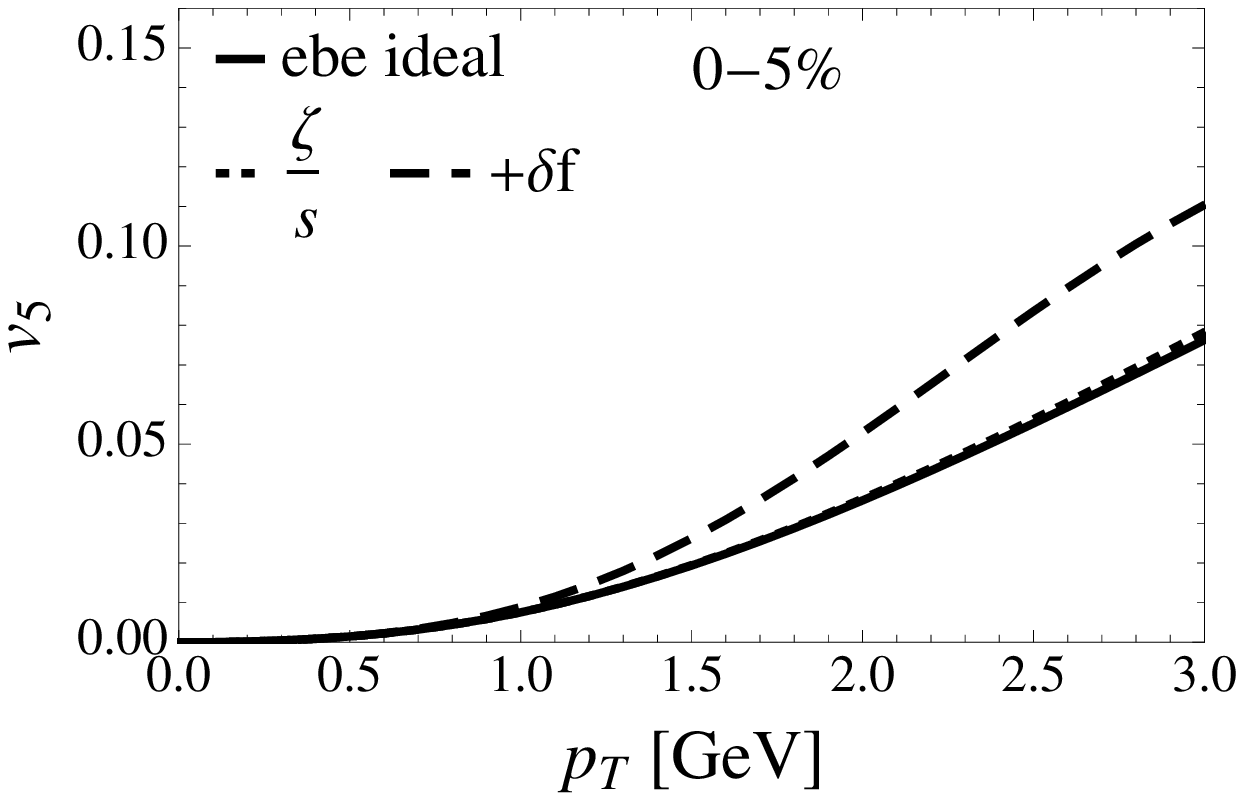} & %
\includegraphics[width=0.4\textwidth]{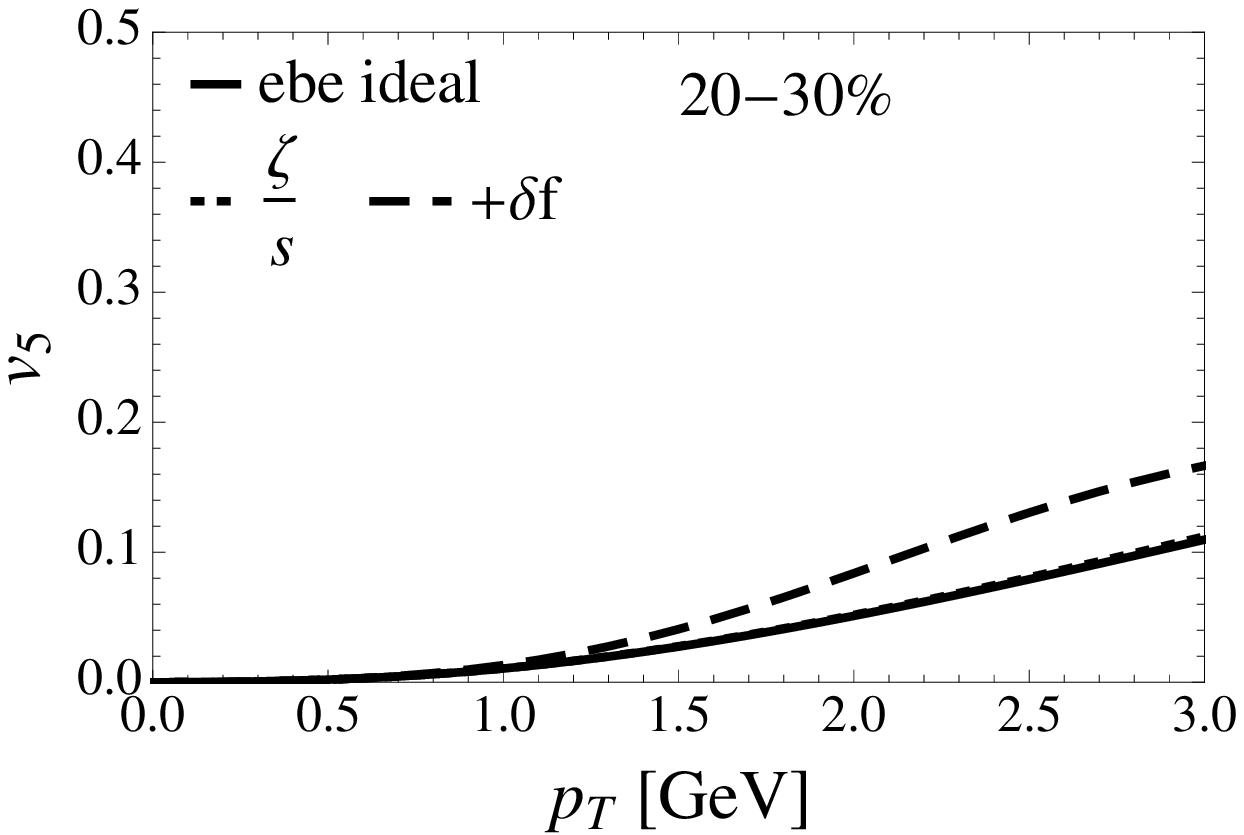}%
\end{tabular}%
\caption{Results for the bulk viscosity $\protect\zeta/s$ shown for most
central collisions ($0-5\%$) and non-central collisions ($20-30\%$) computed
using event-by-event simulations. The solid lines corresponds to the ideal
fluid result, the short dashed lines include bulk viscosity only on the
hydrodynamical evolution but not at freeze-out while the long dashed lines
include bulk viscosity effects both on the hydro evolution and at
freeze-out. }
\label{tab:gt1}
\end{figure}

In Fig.\ \ref{tab:gt1} the flow harmonics from $v_{2}$ to $v_{5}$ are shown
for our two centrality classes for event-by-event Glauber initial
conditions. In all the different plots in Fig.\ \ref{tab:gt1}, solid lines
correspond to the ideal fluid dynamics solution, whereas short dashed curves
include the effects of bulk viscosity only on the hydrodynamical evolution
(i.e., $\delta f$=0 at freeze-out) and the long dashed curve takes into
account the effects of bulk viscosity both in the hydro evolution and at
freeze-out.

Note that, similarly to what is generally seen in the case of shear
viscosity \cite{Dusling:2009df}, when the contribution from $\delta f$ is
not included in the Cooper-Frye procedure the overall effect of bulk
viscosity on the differential flow anisotropies is small, with basically no
deviation from the ideal fluid result. On the other hand, when one considers
the additional non-equilibrium correction $\delta f$ there is a universal
enhancement in the $v_{n}$'s regardless of centrality class, mostly above $%
p_{T}=1$ GeV. The effect is most significant in non-central collisions.
Thus, bulk viscosity affects higher order flow harmonics in the opposite way
that shear viscosity does. In fact, while shear viscosity suppresses $%
v_n(p_T)$, bulk viscosity actually enhances it. One could expect that in the
case where both shear and bulk viscosity are included in event by event
simulations there could be some competition between the two effects. This
interesting question is left for a future study.

\begin{figure}[tbp]
\centering
\begin{tabular}{cc}
\includegraphics[width=0.4\textwidth]{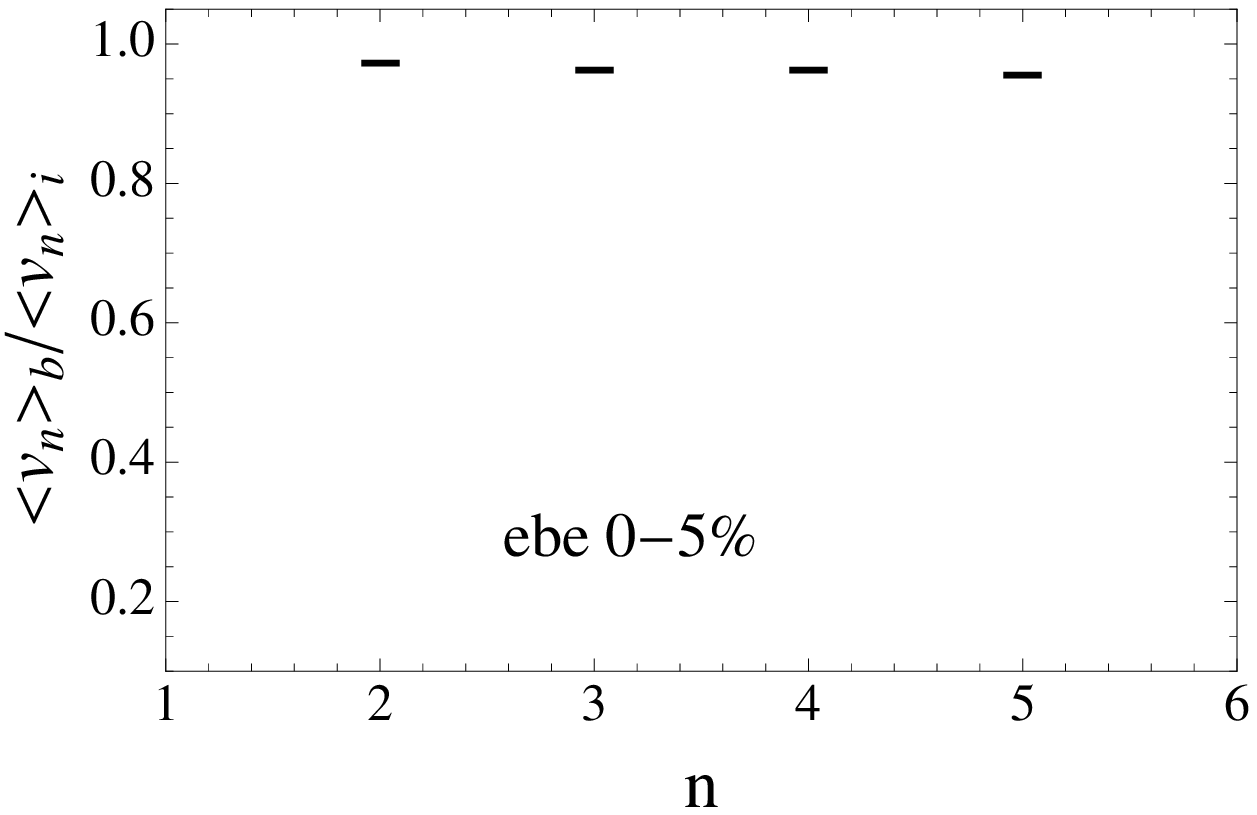} & %
\includegraphics[width=0.4\textwidth]{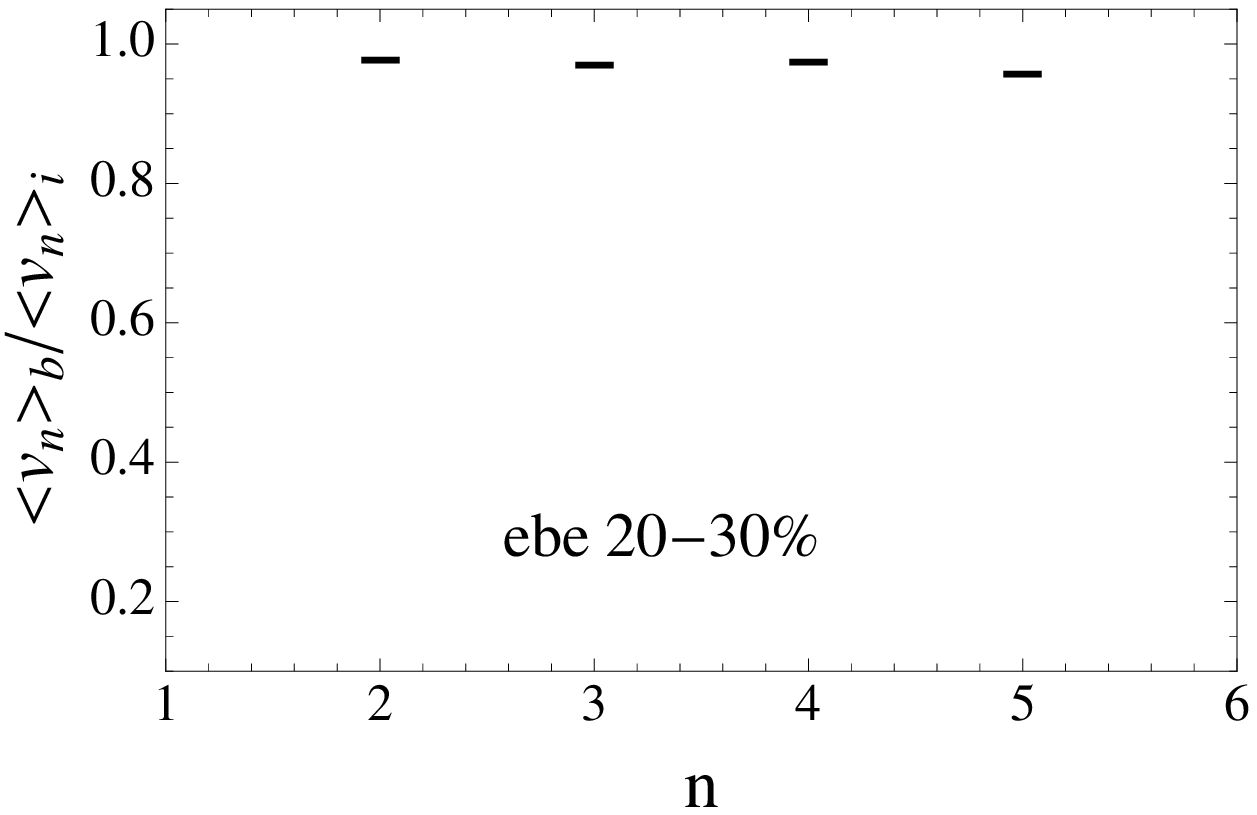} \\ 
& 
\end{tabular}%
\caption{Ratio between the integrated flow coefficients $v_n$'s including
bulk viscosity effects and the corresponding ideal fluid results, computed
in event by event simulations, as a function of $n$ for most central
collisions ($0-5\%$) and peripheral collisions ($20-30\%$). }
\label{tab:vint}
\end{figure}

In Fig.\ \ref{tab:vint} we show our results for the integrated $v_{n}$
coefficients for the two centrality classes. The plot shows $v_{n}$ divided
by the corresponding ideal fluid result as a function of the mode number $n$%
. For the bulk viscosity and relaxation time coefficients used in this work,
we found that the integrated $v_{n}$'s computed in the viscous fluid are
only slightly lower than those found for the ideal fluid, for both
centrality classes. This indicates that the value of $\zeta /s$ chosen in
this work is small and does not affect the fluid 4-velocity and temperature
by much.

\begin{figure}[tbp]
\centering
\begin{tabular}{cc}
\includegraphics[width=0.4\textwidth]{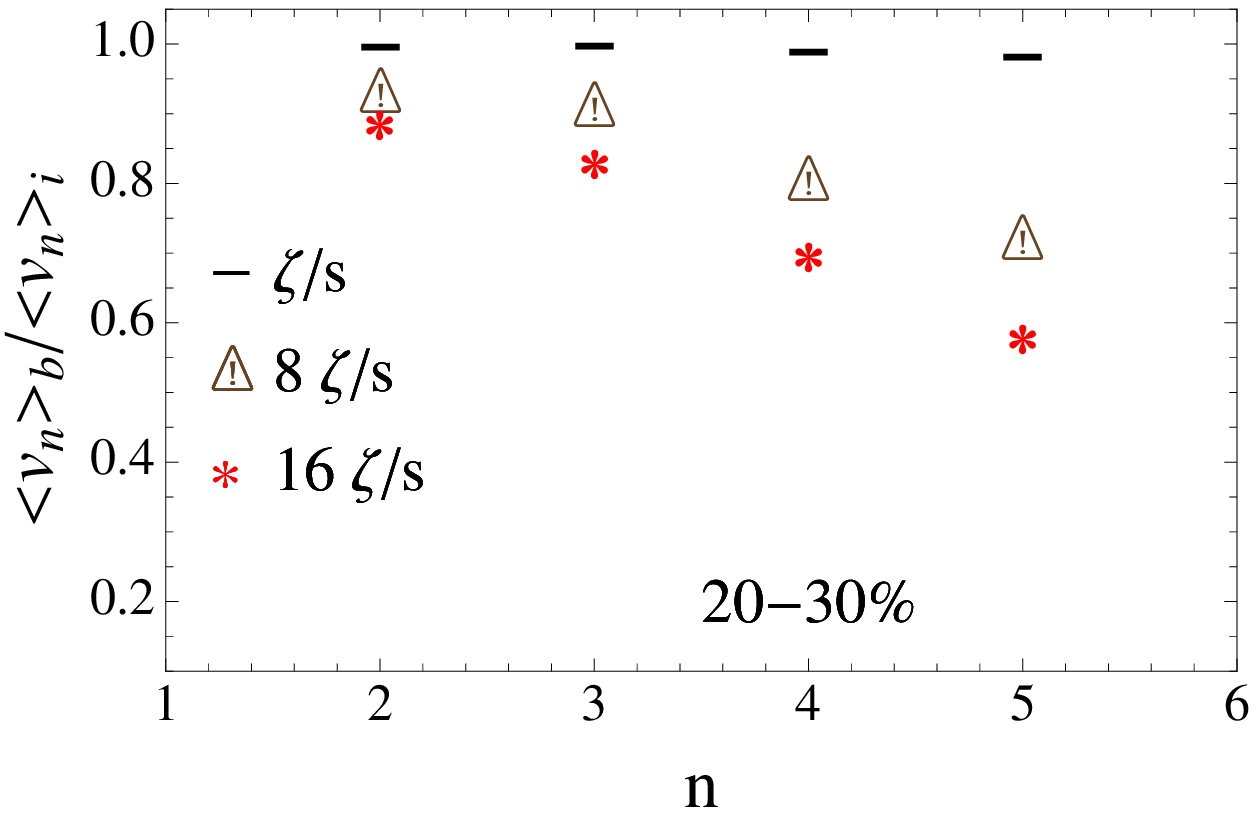} & %
\end{tabular}%
\caption{Ratio between the integrated flow coefficients $v_n$'s including
bulk viscosity effects and the corresponding ideal fluid results as a function of $n$.
This calculation was performed using an initial condition averaged over $150$ MC Glauber events
of the $20$--$30\%$ centrality class.  }
\label{newFig}
\end{figure}

On the other hand, as one increases the bulk viscosity coefficient the
viscous effects on the integrated $v_{n}$ can become large (when compared to
the ideal fluid solution, $v_{n}^{\mathrm{ideal}}$). In Fig.~\ref{newFig} we
show the ratio $v_{n}/v_{n}^{\mathrm{ideal}}$ computed for several values of 
$\zeta /s$. This simulation was performed for an initial condition
constructed from an average over $150$ MC Glauber events taken from the $20$%
--$30\%$ centrality class. Note that, even though this initial condition is
considerably smoother than the usual MC Glauber one, it still has a finite $%
v_{3}$ and $v_{5}$. Also, the integrated $v_{n}$'$s$ showed in Fig.~\ref%
{newFig} were computed without the $\delta f$ correction. We remark that the 
$\delta f$ correction has a very small effect on integrated flow harmonics
and should not contribute much for this plot. One can see that, when the $%
\zeta /s$ taken from Eq.~(\ref{eqn:adszeta}) is multiplied by $8$ (leaving it
with approximately the same magnitude as the shear viscosity coefficient, $%
\zeta /s\sim 0.08$) the flow harmonics are considerably reduced by bulk
viscosity. If we multiply it by $16$, the effect is even greater. This
result indicates that, if the order of magnitude of the bulk viscosity is
close to the one expected for the shear viscosity (as may happen in the
hadronic phase), it is not a good approximation to neglect it.

We did not compute $v_{n}\left( p_{T}\right) $ for such larger values of $%
\zeta /s$ because the $\delta f$ correction (for any of the $\delta f$'s
discussed in this paper) becomes too large and, consequently, renders the
resulting calculation meaningless. This indicates several possibilities: 1)
the bulk viscous $\delta f$'s computed so far in the field are still not
precise enough 2) the bulk viscosity coefficient is actually very small, and/or
3)  the $\delta f$ originating from shear viscosity will cancel the one from
bulk, allowing for larger values of bulk viscosity to be used even for the
currently existing $\delta f$'s. From the results of this paper, we are not
able to state which of the above is actually true.

\section{Conclusions}

In this paper we used the newly developed, Lagrangian 2+1 viscous
hydrodynamic code v-USPhydro to study the effects of bulk viscosity on the
collective flow harmonics observed in ultrarelativistic heavy ion
collisions. We found that flow harmonics can be significantly affected by
bulk viscosity effects even in the case where the maximum of the temperature
dependent $\zeta/s$ is nearly an order of magnitude smaller than the
standard $\eta/s =1/(4\pi)$ value commonly used in hydrodynamical
simulations including only shear viscosity. The inclusion of bulk viscous
effects at freeze-out for a system of multi-hadron species was computed
using the Moments method \cite{Denicol:2012cn,Denicol:2012yr}, which led to
a consistent non-equilibrium correction to the distribution function of
pions that remains well behaved when $p_T =0-3$ GeV. It would be interesting
to investigate how the inclusion of heavier hadrons and experimentally
measured hadron cross sections affects the coefficients in $\delta f$ of
each hadron species.

We performed event by event simulations (using Monte-Carlo Glauber initial
conditions) that allowed us to study for the first time the effects of bulk
viscosity on collective flow harmonics of higher order. We have found that
bulk viscosity enhances the differential flow coefficients $v_{n}(p_{T})$,
for $n=$ 2, 3, 4, and 5, with respect to their ideal fluid values when $%
p_{T}\sim 1-3$ GeV. This shows that bulk viscosity affects differential flow
anisotropies in the opposite way than that found in the case of shear
viscosity, which is known to lead to an overall suppression of $v_{n}(p_{T})$
in the same intermediate $p_{T}$ range. Thus, our results indicate that a
realistic description of the QGP hydrodynamical evolution should include
(preferably temperature dependent) shear and bulk viscosities in order to
correctly describe the suppression of differential flow harmonics within
relativistic heavy ion collisions. The bulk viscosity driven enhancement of $%
v_{n}(p_{T})$ found in this paper also opens up the interesting possibility
that bulk and shear viscosity effects may actually compete in the
suppression of flow anisotropies in the viscous QGP.

It should be noted that all the $\delta f$'s used in this work imply that
even a small bulk viscosity can have a large effect on differential flow
harmonics. This happens for values of bulk viscosity that do not even affect
the fluid-dynamical evolution of the plasma. For values of bulk viscosity of 
$\zeta /s\sim 0.08$, which actually have a considerable effect on the
fluid-dynamical evolution of the system, the $\delta f$ correction arising
from bulk terms becomes too large, making it physically meaningless to
apply it to compute $p_{T}$--differential observables. Therefore, it is
important to verify how precise the current $\delta f$'s in the field
actually are. For the case of the moment expansion this can be verified by
checking the convergence of the series. It is possible that the truncation
employed in this work is still far from the converged solution, but this can
only be confirmed in a future work.  

J.~Noronha-Hostler, R.~P.~G.~Andrade, J.~Noronha, and F.~Grassi acknowledge
Funda\c{c}\~{a}o de Amparo \`{a} Pesquisa do Estado de S\~{a}o Paulo
(FAPESP) and Conselho Nacional de Desenvolvimento Cient\'{\i}fico e Tecnol%
\'{o}gico (CNPq) for financial support. G.~S.~Denicol is supported by the
Natural Sciences and Engineering Research Council of Canada. The authors
thank T.~Kodama, Y.~Hama, and F.~Gardim for discussions on the SPH approach
to relativistic fluid dynamics and A.~Dumitru for providing the Monte Carlo
Glauber model.

\clearpage
\appendix

\section{Relativistic Fluid Dynamic Equations in the SPH Formalism}

\label{appendixSPH}

In the SPH approach one introduces a conserved reference density current $%
J^{\mu }=\sigma u^{\mu }$ where $\sigma $ is the local density of a fluid
element in its rest frame. As the fluid flows, the cell is deformed but its
density obeys $D\sigma +\sigma \theta =0$, which in hyperbolic coordinates
is equivalent to $\partial _{\mu }(\tau \sigma u^{\mu })=0$. In terms of
this reference density, the equations of motion used in this paper can be
written as \cite{Denicol:2009am} 
\begin{eqnarray}
\gamma \frac{d}{d\tau }\left[ \frac{\left( \varepsilon +p+\Pi \right) }{%
\sigma }u^{\mu }\right] &=&\frac{1}{\sigma }\partial ^{\mu }\left( p+\Pi
\right)  \label{eqboa1} \\
\gamma \frac{d}{d\tau }\left( \frac{s}{\sigma }\right) +\left( \frac{\Pi }{%
\sigma }\right) \frac{\theta }{T} &=&0  \label{eqboa2} \\
\tau _{\Pi }\gamma \frac{d}{d\tau }\left( \frac{\Pi }{\sigma }\right) +\frac{%
\Pi }{\sigma }+\left( \frac{\zeta }{\sigma }\right) \theta &=&0\,.
\label{eqboa3}
\end{eqnarray}

These equations are completely equivalent to those in Section \ref{eom} but
they are more suitable for the Lagrangian implementation via SPH, as will be
explained in the following.

\begin{figure}[tbp]
\includegraphics[width=0.4\textwidth]{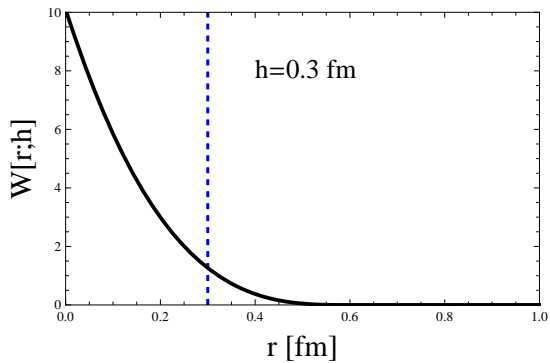}
\caption{Kernel function $W[\mathbf{r};h]$ as a solid black curve for $h=0.3$
fm (denoted by the vertical dashed blue line) used in this paper.}
\label{kernel}
\end{figure}

The fundamental idea behind a mesh free method such as SPH is that the boost
invariant hydrodynamical fields can be reconstructed, at any given point in
space and time, using a discrete set of Lagrangian coordinates $\left\{%
\mathbf{r}_{\alpha }(\tau),\alpha =1,...,N_{SPH}\right\}$ together with a
normalized piece-wise distribution function $W\left[\mathbf{r};h\right]$ for
the discretization procedure. The kernel is chosen to have a finite support
given by $h$, i.e, its value strictly vanishes for $|\mathbf{r}| \gg h$.
Also, the kernel is a delta sequence in the sense that $\lim_{h\to 0}W[%
\mathbf{r};h]=\delta(\mathbf{r})$. The parameter $h$ is a length scale that
represents the width of the kernel and it defines a cutoff for modes with
shorter wavelength. The smaller the $h$ the larger is the number of SPH
particles needed to accurately describe the flow. In general, the choice of $%
h$ dictates how much of the initial structure in the initial conditions will
be reproduced and used as initial values for the subsequent dynamics. In
practice, the actual size of $h$ is also limited by the computational time
available. We shall discuss our choice for $h$ in more detail in the next
section. For boost invariant hydrodynamics, the kernel function (in
hyperbolic coordinates) is normalized as 
\begin{eqnarray}
\int W\left[\mathbf{r};h\right] d^2\mathbf{r} &=&1\,.
\end{eqnarray}
The Kernel function used in v-USPhydro can be seen in Fig.\ \ref{kernel}.

As was mentioned above, the conserved reference current density obeys the
equation $\partial_\mu(\tau \sigma u^\mu)=0$ in hyperbolic coordinates.
Within SPH, the reference density in the lab frame is expressed in
Lagrangian coordinates as 
\begin{equation}
\tau \gamma \sigma \rightarrow \sigma^{*}\left(\mathbf{r},\tau\right)=\sum_{%
\alpha =1}^{N_{SPH}}\nu_{\alpha}\,W\left[\mathbf{r}-\mathbf{r}_\alpha(\tau);h%
\right]  \label{sigma}
\end{equation}
where $\nu_\alpha$ are constants. Due to the normalization of the kernel,
one can see that integral of the reference density in the transverse plane
is a constant, i.e., $\int d^2\mathbf{r}\, \sigma^{*}\left(\mathbf{r}%
,\tau\right)=\sum_{\alpha=1}^{N_{SPH}}\nu_\alpha$. Therefore, it is natural
to interpret the quantity $\nu_\alpha$ as a conserved quantity attached to
the Lagrangian coordinate $\mathbf{r}_\alpha(\tau)$ and $\sigma^{*}$ as a
sum of small piece-wise distributions $\nu_\alpha W[\mathbf{r}-\mathbf{r}%
_\alpha(\tau);h]$, which are called ``SPH particles".

One now defines the vector current 
\begin{gather}
\mathbf{j}^*\left( \mathbf{r},\tau\right) =\sum_{\alpha =1}^{N_{SPH}}\nu
_{\alpha }\frac{d\mathbf{r}_{\alpha }(\tau)}{d\tau}W[\mathbf{r}-\mathbf{r}%
_{\alpha }(\tau);h],  \label{current}
\end{gather}%
so that the continuity equation $\partial_\tau \sigma^{*}\left(\mathbf{r}%
,\tau\right)+\nabla_{\mathbf{r}} \cdot \mathbf{j}^*\left( \mathbf{r}%
,\tau\right)=0$ for the reference density is automatically satisfied \cite%
{Aguiar:2000hw}. Eq.\ (\ref{current}) shows that each Lagrangian coordinate
(or SPH particle) $\mathbf{r}_{\alpha }(\tau)$ has velocity $\mathbf{u}%
_\alpha(\tau)=\gamma_\alpha(\tau)\mathbf{v}_{\alpha }(\tau)$, where $\mathbf{%
v}_{\alpha }(\tau)=d\mathbf{r}_{\alpha }(\tau)/d\tau$ and $\gamma_\alpha=1/%
\sqrt{1-\mathbf{v}_\alpha^2}$, and it carries a quantity $\nu _{\alpha }$
for the reference density $\sigma^{\ast }$.

Now, let $a(\mathbf{r},\tau)$ be the density associated with some extensive
quantity. The SPH description of this quantity is 
\begin{equation}
a(\mathbf{r},\tau)=\sum_{\alpha =1}^{N_{SPH}}\nu_{\alpha}\,\frac{a(\mathbf{r}%
_\alpha(\tau))}{\sigma^{*}\left(\mathbf{r}_\alpha(\tau)\right)}W\left[%
\mathbf{r}-\mathbf{r}_\alpha(\tau);h\right]\,.
\end{equation}
One can then see that any spatial gradient of $a(\mathbf{r},\tau)$ acts only
on the kernel function and this gradient is still a smooth function. For
instance, for the zeroth component of the entropy current in the lab frame $%
s^*=s \gamma \tau$ one finds 
\begin{equation}
s^*(\mathbf{r},\tau)=\sum_{\alpha =1}^{N_{SPH}}\nu_{\alpha}\,\frac{s(\mathbf{%
r}_\alpha(\tau))}{\sigma\left(\mathbf{r}_\alpha(\tau)\right)}W\left[\mathbf{r%
}-\mathbf{r}_\alpha(\tau);h\right]\,
\end{equation}
while for the bulk term 
\begin{equation}
\Pi(\mathbf{r},\tau)=\sum_{\alpha=1}^{N_{SPH}}\nu _{\alpha }\frac{1}{\gamma
_{\alpha }\tau }\left( \frac{\Pi }{\sigma }\right) _{\alpha }W[\mathbf{r}-%
\mathbf{r}_{\alpha }(\tau);h]\,.
\end{equation}

The dynamical variables in the SPH method are then 
\begin{equation}
\left\{ \mathbf{r}_{\alpha },\mathbf{u}_{\alpha },\left( \frac{s}{\sigma }%
\right) _{\alpha },\left( \frac{\Pi }{\sigma }\right) _{\alpha };\ \alpha
=1,..,N_{SPH}\right\}  \label{dynamicalvariables}
\end{equation}%
and they represent the position, velocity, entropy, and bulk viscosity
associated with the $\alpha $-th SPH particle, respectively. Using that $%
\partial_\mu(\tau \sigma u^\mu)=0$, we see that $\sigma^*\nabla \cdot 
\mathbf{v}=-\frac{d\sigma^*}{d\tau}$ and, consequently, the fluid expansion
rate for each SPH particle is 
\begin{equation}
\theta_\alpha = (D_\mu u^\mu)_\alpha=\frac{d\gamma_\alpha}{d\tau}+\frac{%
\gamma_\alpha}{\tau}-\frac{\gamma_\alpha}{\sigma_\alpha^*}\frac{%
d\sigma_\alpha^*}{d\tau}\,.
\end{equation}

From the equations of motion in (\ref{eqboa1}-\ref{eqboa3}) we obtain the
following equations associated with each SPH particle 
\begin{eqnarray}
\sigma^{\ast }\frac{d}{d\tau }\left( \frac{\left( \varepsilon +p+\Pi \right)
_{\alpha }}{\sigma _{\alpha }}\,u_{i\ \alpha }\right) &=&\tau \sum_{\beta
=1}^{N_{SPH}}\nu _{\beta }\sigma _{\alpha }^{\ast }\left( \frac{p_{\beta
}+\Pi _{\beta }}{\left( \sigma _{\beta }^{\ast }\right) ^{2}}+\frac{%
p_{\alpha }+\Pi _{\alpha }}{\left( \sigma _{\alpha }^{\ast }\right) ^{2}}%
\right) \ \partial _{i}W[\mathbf{r}_{\alpha }-\mathbf{r}_{\beta }(\tau);h]\,,
\label{MotionSPH}
\end{eqnarray}%
\begin{equation}
\gamma_\alpha\frac{d}{d\tau}\left(\frac{s}{\sigma}\right)_\alpha+\left(\frac{%
\Pi}{\sigma}\right)_\alpha\left(\frac{\theta}{T}\right)_\alpha=0\,,
\label{entropyequationSPH}
\end{equation}
and 
\begin{equation}
\tau _{\Pi_{\alpha }}\gamma _{\alpha }\frac{d}{d\tau }\left( \frac{\Pi }{%
\sigma }\right) _{\alpha }+\left( \frac{\Pi }{\sigma }\right) _{\alpha
}+\left(\frac{\zeta}{\sigma}\right)_{\alpha }\theta _{\alpha }=0\,.
\label{PiequationSPH}
\end{equation}

The r.h.s. of Eq.\ (\ref{MotionSPH}) is the SPH representation of the
gradients of pressure and bulk viscosity and, in the case of vanishing bulk
viscosity, these equations become those found using the variational
principle \cite{Aguiar:2000hw}. Eqs.\ (\ref{MotionSPH}-\ref{PiequationSPH})
are the SPH representation of the equations of motion which are solved in
v-USPhydro. The beauty of the SPH method to solving hydrodynamics is that
the coupled, nonlinear partial differential equations in the Eulerian view
are described in terms of a set of nonlinear ordinary coupled differential
equations for the Lagrangian variables.

We remark that in our Lagrangian approach no numerical viscosity is needed
even in the inviscid case. Moreover, in our approach no extra conditions on
the dynamical fields at very low temperatures needs to be imposed. In fact,
we show below that our code in the absence of bulk viscosity exactly matches
the analytical solution for 2+1 inviscid conformal hydrodynamics derived by
in \cite{Gubser:2010ze}, which provides a very stringent test of our
approach to solve the equations of relativistic fluid dynamics. We have also
checked that our code matches the calculations performed in \cite%
{Denicol:2009am}.

\subsection*{Comparison to Gubser flow}

\label{Gubserflow}

In \cite{Gubser:2010ze}, an analytical solution of 2+1 (i.e., boost
invariant) ideal conformal (i.e., $\varepsilon=3p$) hydrodynamics was
derived that can be used as a nontrivial check for the numerical
hydrodynamic codes. They found the following analytical solution for the
energy density profile 
\begin{equation}  \label{eqn:epsilon_GT}
\varepsilon(\tau,r)=\frac{\varepsilon_0}{\tau^{4/3}}\frac{(2q)^{8/3}}{\left[%
1+2q^2\left(\tau^2+r^2\right)+q^4\left(\tau^2-r^2\right)^2\right]^{4/3}}
\end{equation}
where 
\begin{equation}
r^2=x^2+y^2
\end{equation}
and $q$ (in 1/fm) and $\varepsilon_0$ are constants set to 1 (the overall
magnitude of the energy density, set by $\varepsilon_0$, is immaterial for
this type of check). The analytical solution for the flow is \cite%
{Gubser:2010ze} 
\begin{eqnarray}
u_x(\tau,r)&=&\frac{\sinh \left[\kappa (\tau,r)\right]x}{r}  \nonumber \\
u_y(\tau,r)&=&\frac{\sinh \left[\kappa (\tau,r)\right]y}{r}
\label{eq:radialV}
\end{eqnarray}
where 
\begin{equation}
\kappa (\tau,r)=\mathrm{arctanh}\left(\frac{2q^2\tau r}{1+q^2 \tau^2+q^2 r^2}%
\right)\,.
\end{equation}

\begin{figure}[tbp]
\centering
\epsfig{file=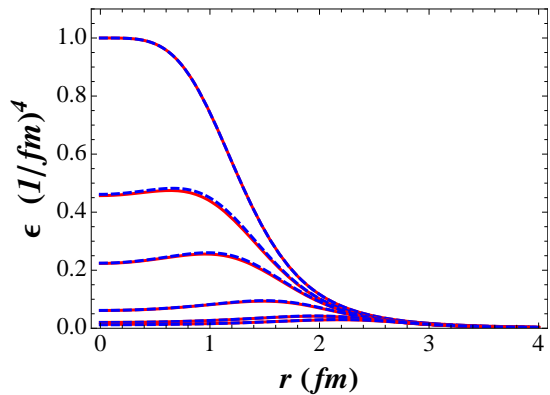,width=0.4\linewidth,clip=}
\caption{Comparison between our numerical results for the energy density
computed using v-USPhydro (dashed blue lines) and the analytical solution in
Eq.\ (\protect\ref{eqn:epsilon_GT}), shown as solid red lines, for times $%
\protect\tau$ = 1, 1.2, 1.4, 1.8, 2.2, and 2.4 fm. The energy density at the
origin starts at $1$ fm$^{-4}$ at $\protect\tau_0=1$ fm and is reduced by an
order of magnitude after 1.4 fm. }
\label{fig:e_GT}
\end{figure}

\begin{figure}[tbp]
\centering
\epsfig{file=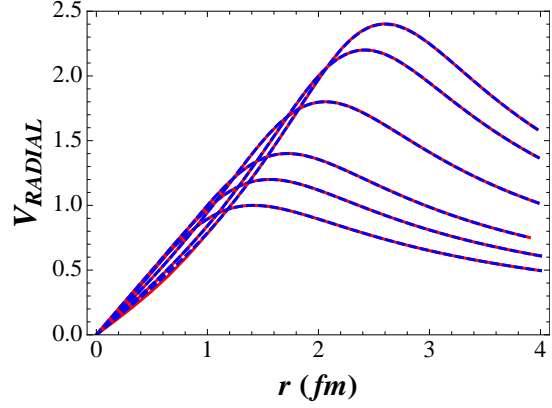,width=0.4\linewidth,clip=}
\caption{Comparison between our numerical results for the radial velocity $%
\protect\sqrt{u_x^2+u_y^2}$ computed using v-USPhydro (dashed blue lines)
and the analytical solution in Eq.\ (\protect\ref{eq:radialV}), shown as
solid red lines, for times $\protect\tau$ = 1, 1.2, 1.4, 1.8, 2.2, and 2.4
fm. The initial transverse flow has a peak around 1 at $r\sim 1.4$ fm and
peaks around 2.5 at $r\sim 2.6$ fm when $\protect\tau=2.4$ fm.}
\label{fig:rv_GT}
\end{figure}

We take the analytical formulas shown above computed at $\tau_0$ =1 fm to
define the initial conditions for the energy density and flow. These initial
conditions are used as input for our numerical code and we then compare our
numerical results to the analytical solution for several values of $\tau$,
as shown in Figs.\ \ref{fig:e_GT}-\ref{fig:rv_GT}. 

This system expands very rapidly, the energy density falls very steeply, and
it requires a relatively small $h$, which in turn requires a great deal of
SPH particles to achieve an accurate description of the analytical solution.
v-USPhydro was able to match the analytical solution very well using $h=0.12$
fm and a total of $103041$ SPH particles, as seen in Figs.\ \ref{fig:e_GT}-%
\ref{fig:rv_GT}. Note, however, that the Glauber initial conditions used in
the numerical analysis in this paper do not expand as fast as this
analytical solution (they also do not have initial transverse flow) and not
as many SPH particles are needed in this case. The dependence of our results
for the flow coefficients shown in Section \ref{results} with the choice of $%
h$ and the total number of SPH particles is studied in Appendix \ref%
{sec:convh}.

\section{Some Details about the Cooper-Frye Freeze-Out within the SPH
Approach}

\label{cooperfryeappendix}

In this section we present some details about the Cooper-Frye formalism
employed in hadronic freeze-out \cite{Cooper:1974mv} within the SPH approach 
\cite{Osada:2001hw,Hama:2004rr,denicoltesemestrado}. In the Cooper-Frye
freeze-out procedure \cite{Cooper:1974mv}, the particle flux through an
isothermal hypersurface $\Sigma$ defines the momentum distribution of a
given particle species of degeneracy $g$, mass $m$, energy $E=\sqrt{p_T^2+m^2%
}$, and distribution function $f$ 
\begin{equation}
\frac{dN}{dyp_T dp_Td\phi} = \frac{g}{(2\pi)^3}\int_{\Sigma}d\Sigma \cdot
p\, \,f(p\cdot u,\Pi,T_{FO}) \,.
\end{equation}
where $T_{FO}$ is the freeze-out temperature, $y$ is the particle's
rapidity, and $\Pi$ is the bulk viscosity contribution. Note that $p\cdot u$
and $\Pi$ are the only Lorentz invariant structures that need to be taken
into account to describe the viscous effects coming solely from bulk
viscosity.

As explained in \cite{Osada:2001hw,Hama:2004rr,denicoltesemestrado}, in the
SPH formalism the integral over the isothermal hypersurface is written in
terms of a sum of SPH particles as 
\begin{equation}
\frac{dN}{dyp_T dp_Td\phi} = \frac{g}{(2\pi)^3}\,\sum_{\alpha=1}^{N_{SPH}}%
\frac{(p\cdot n)_\alpha}{ (n\cdot u)_\alpha }\frac{\nu_\alpha}{\sigma_\alpha}
\,f(T_{FO}, (p\cdot u)_\alpha,\Pi_\alpha)
\end{equation}
where the index $\alpha$ indicates the SPH particle, $N_{SPH}$ is the total
number of SPH particles, $(n_\mu)_\alpha$ is the normal vector of the
isothermal hypersurface reconstructed using the $\alpha$-th SPH particle, $%
(u_\mu)_\alpha$ is the 4-velocity of the SPH particle, and $\Pi_\alpha$ is
the bulk viscosity of the SPH particle. The distribution function in the
equation above is the sum of the ideal distribution function and the
non-equilibrium correction $\delta f$ shown in Eq.\ (\ref{bulkmoments}) (for
the sake of generality, here we do not use classical statistics).

In explicit form, the particle distribution for a given particle species is 
\begin{equation}
\frac{dN}{dyp_Tdp_Td\phi} = \frac{g}{(2\pi)^3}\,\sum_{\alpha=1}^{N_{SPH}}\,%
\left[ q_{0 \,\alpha} \,\mathcal{I}_1(\alpha,m,T_{FO})-(\mathbf{p}_T \cdot 
\mathbf{q}_{T})_\alpha\, \mathcal{I}_2(\alpha,m,T_{FO}) \right]\,.
\end{equation}
where 
\begin{equation}
(q_\nu)_\alpha= \frac{(n_\nu)_\alpha}{| (n \cdot u)_\alpha |}\frac{\nu_\alpha%
}{\sigma_\alpha}
\end{equation}
and 
\begin{eqnarray}
\mathcal{I}_1(\alpha,m,T_{FO})&=& 2\sum_{n=0}^{\infty} (-a)^n
\lambda^{n+1}_\alpha\left\{E^2 F_\alpha^{(1)}(n) K_0\left[\frac{(n+1)E
\gamma_\alpha}{T_{FO}}\right] \right. \\
&+& \left. K_1\left[\frac{(n+1)E \gamma_\alpha}{T_{FO}}\right] \,E \left(
F_\alpha^{(0)}(n)+\frac{F_\alpha^{(1)}(n)T_{FO}}{(n+1)\gamma_\alpha}+E^2
F_\alpha^{(2)}(n) \right) +\frac{E^2 F_\alpha^{(2)}(n)T_{FO}}{%
(n+1)\gamma_\alpha} K_2\left[\frac{(n+1)E \gamma_\alpha}{T_{FO}}\right]%
\right \}  \nonumber
\end{eqnarray}
and 
\begin{eqnarray}
\mathcal{I}_2(\alpha,m,T_{FO})&=& 2\sum_{n=0}^{\infty} (-a)^n
\lambda^{n+1}_\alpha\left\{ K_0\left[\frac{(n+1)E \gamma_\alpha}{T_{FO}}%
\right] (F_\alpha^{(0)}(n)+F_\alpha^{(2)}(n)E^2)\right. \\
&+& \left. E K_1\left[\frac{(n+1)E \gamma_\alpha}{T_{FO}}\right]%
\left(F_\alpha^{(1)}(n)+\frac{F_\alpha^{(2)}(n)T_{FO}}{(n+1)\gamma_\alpha}%
\right)\right \}  \nonumber
\end{eqnarray}
with $a=1\left( -1\right) $ for fermions(bosons) and $0$ for classical
particles, $\lambda_\alpha = e^{\,(\mathbf{p}_T\cdot \mathbf{u}%
_T)_\alpha/T_{FO}}$, $K_n[x]$ is a modified Bessel function, and 
\begin{equation}
F_\alpha^{(0)}(n)=1+(n+1)\Pi_\alpha^{pion}\left[B_0 -D_0 (\mathbf{p}_T\cdot 
\mathbf{u}_T)_\alpha+ E_0 (\mathbf{p}_T\cdot \mathbf{u}_T)_\alpha^2\right]\,,
\end{equation}
\begin{equation}
F_\alpha^{(1)}(n)=(n+1)\Pi_\alpha\gamma_\alpha\left[D_0 -2 E_0 (\mathbf{p}%
_T\cdot \mathbf{u}_T)_\alpha\right]\,,
\end{equation}
\begin{equation}
F_\alpha^{(2)}(n)=(n+1)\Pi_\alpha\gamma_\alpha^2\,E_0\,.
\end{equation}

\section{Event Plane Method for Collective Flow Coefficients}

\label{eventplanemethod}

We use the event plane method to compute the collective flow coefficients 
\cite{Poskanzer:1998yz}. First, we use that given the differential number $%
\frac{dN_i(p_T,\phi)}{dyp_T dp_T d\phi}$ of hadrons of species $i$ in a
given event, we can integrate it over the azimuthal angle $\phi$ to find the
spectrum 
\begin{equation}
\frac{dN_i(p_T)}{dyp_T dp_T}=\int_0^{2\pi} d\phi \frac{dN_i}{dyp_T dp_T d\phi%
}\,.
\end{equation}
We then define the event plane vectors 
\begin{equation}
Q_x^i[n] = \int_{p_{T\,min}}^{p_{T\,max}}dp_T\, p_T^2 \int_0^{2\pi}d\phi
\,\cos(n\phi) \frac{dN_i}{dyp_T dp_T d\phi}\,,
\end{equation}
\begin{equation}
Q_y^i[n] = \int_{p_{T\,min}}^{p_{T\,max}}dp_T\, p_T^2 \int_0^{2\pi}d\phi
\,\sin(n\phi) \frac{dN_i}{dyp_T dp_T d\phi}
\end{equation}
and the event plane angles 
\begin{equation}
\psi^i[n] =\frac{1}{n} \tan^{-1}\left(\frac{Q^i_y[n]}{Q^i_x[n]}\right)\,,
\end{equation}
where in this paper $p_{T\,min}=0$ GeV and $p_{T\, max}=3$ GeV.

The collective flow coefficients as functions of the transverse momentum are
then 
\begin{equation}
v_n^i(p_T)=\frac{\int_0^{2\pi}d\phi \,\frac{dN_i}{dyp_T dp_T d\phi} \cos[{{%
n(\phi-\psi^i[n])}}]}{\frac{dN_i(p_T)}{dyp_T dp_T}}\,.
\end{equation}
The integrated $v_n$'s are given by 
\begin{equation}
v_n^i = \frac{\int_{p_{T\,min}}^{p_{T\,max}}dp_T \; p_T \int_0^{2\pi}d\phi \,%
\frac{dN_i}{dyp_T dp_T d\phi} \cos[{{n(\phi-\psi^i[n])}}]}{%
\int_{p_{T\,min}}^{p_{T\,max}}dp_T \; p_T \frac{dN_i(p_T)}{ dyp_T dp_T}}\,.
\end{equation}

\section{Formulas for the Coefficients in $\protect\delta f$}

\label{Deltaf_Coeffs}

In this paper, $\delta f_{\mathbf{k}}^{\left( i\right) }$ is computed
following the steps outlined in Refs.\ \cite{Denicol:2012cn,Denicol:2012yr}.
In this Appendix, we just outline how the coefficients that appear in $%
\delta f_{\mathbf{k}}^{\left( i\right) }$ can be computed.

The coefficient $a_{nr}^{(0)i}$ is a function of the temperature and the
mass of the $i$--th hadron species,

\begin{eqnarray*}
\frac{a_{10}^{\left( 0\right) i}}{a_{11}^{\left( 0\right) i}} &=&-\frac{%
J_{10}^{i}}{J_{00}^{i}}, \\
\left( a_{11}^{\left( 0\right) i}\right) ^{2} &=&\frac{\left(
J_{00}^{i}\right) ^{2}}{J_{20}^{i}J_{00}^{i}-\left( J_{10}^{i}\right) ^{2}},
\\
\frac{a_{21}^{(0)i}}{a_{22}^{(0)i}} &=&\frac{%
J_{20}^{i}J_{10}^{i}-J_{30}^{i}J_{00}^{i}}{J_{20}^{i}J_{00}^{i}-\left(
J_{10}^{i}\right) ^{2}}, \\
\frac{a_{20}^{(0)i}}{a_{22}^{(0)i}} &=&\frac{J_{10}^{i}J_{30}^{i}-\left(
J_{20}^{i}\right) ^{2}}{J_{00}^{i}J_{20}^{i}-\left( J_{10}^{i}\right) ^{2}},
\\
\left( a_{22}^{(0)i}\right) ^{2} &=&J_{00}^{i}\left( J_{40}^{i}-\frac{\left(
J_{20}^{i}\right) ^{3}-2J_{30}^{i}J_{20}^{i}J_{10}^{i}+\left(
J_{30}^{i}\right) ^{2}J_{00}^{i}}{J_{20}^{i}J_{00}^{i}-\left(
J_{10}^{i}\right) ^{2}}\right) ^{-1}.
\end{eqnarray*}%
The thermodynamic functions $J_{nq}^{i}$ are defined as%
\[
J_{nq}^{i}=g_{i}\int \frac{d^{3}\mathbf{k}}{\left( 2\pi \right) ^{3}k_{i}^{0}%
}\left( u \cdot k_{i}\right) ^{n-2q}\left( -\Delta _{\alpha \beta
}k_{i}^{\alpha }k_{i}^{\beta }\right) ^{q}f_{0\mathbf{k}}^{i}\left( 1-af_{0%
\mathbf{k}}^{i}\right) ,
\]%
where $g_{i}$ is the degeneracy factor, $dK_{i}\equiv $ $d^{3}\mathbf{k/}%
\left[ \left( 2\pi \right) ^{3}k_{i}^{0}\right] $,\ and $f_{0\mathbf{k}}^{i}$
is the local equilibrium distribution function%
\[
f_{0\mathbf{k}}^{i}=\frac{1}{\exp \left( \beta u \cdot k_{i}\right) +a},
\]%
with $a$ being $a=1\left( -1\right) $ for fermions(bosons), and $0$ for
classical particles, and we assumed that the chemical potential is zero. For
completeness, we shall also define the thermodynamic function $I_{nq}^{i}$ 
\[
I_{nq}^{i}=g_{i}\int \frac{d^{3}\mathbf{k}}{\left( 2\pi \right) ^{3}k_{i}^{0}%
}\left( E_{i\mathbf{k}}\right) ^{n-2q}\left( -\Delta _{\alpha \beta
}k_{i}^{\alpha }k_{i}^{\beta }\right) ^{q}f_{0\mathbf{k}}^{i},
\]%
where we defined $E_{i\mathbf{k}}\equiv u \cdot k_{i}$. In this paper we
only consider classical particles for constructing an approximate expression
for the $\delta f_{\mathbf{k}}^{i}$ of hadrons. Therefore, the functions $%
I_{nq}^{i}$ and $J_{nq}^{i}$ are equivalent.

The coefficients $\alpha _{i,m}$ are given by%
\[
\left( 
\begin{array}{c}
\alpha _{i,0} \\ 
\alpha _{i,2}%
\end{array}%
\right) =\mathcal{\hat{M}}_{\left( 0\right) }^{-1}\left( 
\begin{array}{c}
\beta _{i,0} \\ 
\beta _{i,2}%
\end{array}%
\right) ,
\]%
where%
\begin{eqnarray*}
\beta _{i,r} &=&J_{r0}^{i}\frac{J_{30}n_{0}-J_{20}\left( \varepsilon
_{0}+P_{0}\right) }{J_{30}J_{10}-J_{20}J_{20}}\theta -J_{r+1,0}^{i}\frac{%
J_{20}n_{0}-J_{10}\left( \varepsilon _{0}+P_{0}\right) }{%
J_{30}J_{10}-J_{20}J_{20}}\theta -\left[ \left( r-1\right)
I_{r1}^{i}+I_{r0}^{i}\right] \theta \text{ }, \\
J_{nq} &=&\sum_{i=1}^{N}J_{nq}^{i},\text{ }I_{nq}=\sum_{i=1}^{N}I_{nq}^{i}%
\text{.}
\end{eqnarray*}%
Above, $\mathcal{\hat{M}}_{\left( 0\right) }$ is a $\left( 2N-1\right)
\times \left( 2N-1\right) $ matrix that can be derived from the collision
term of the Boltzmann equation. Since we consider only elastic 2-to-2
collisions, it can be simplified to be%
\[
\mathcal{\hat{M}}_{\left( 0\right) }\equiv \left( 
\begin{array}{cc}
\left( 
\begin{array}{ccc}
\mathcal{M}_{11\left( 0\right) }^{00} & \ldots & \mathcal{M}_{1N\left(
0\right) }^{00} \\ 
\vdots & \ddots & \vdots \\ 
\mathcal{M}_{N1\left( 0\right) }^{00} & \ldots & \mathcal{M}_{NN\left(
0\right) }^{00}%
\end{array}%
\right) & \left( 
\begin{array}{ccc}
\mathcal{M}_{11\left( 0\right) }^{02}-\mathcal{M}_{NN\left( 0\right) }^{02}
& \ldots & \mathcal{M}_{1,N-1\left( 0\right) }^{02}-\mathcal{M}_{NN\left(
0\right) }^{02} \\ 
\vdots & \ddots & \vdots \\ 
\mathcal{M}_{N,1\left( 0\right) }^{02}-\mathcal{M}_{NN\left( 0\right) }^{02}
& \ldots & \mathcal{M}_{N,N-1\left( 0\right) }^{02}-\mathcal{M}_{NN\left(
0\right) }^{02}%
\end{array}%
\right) \\ 
\left( 
\begin{array}{ccc}
\mathcal{M}_{11\left( 0\right) }^{20} & \ldots & \mathcal{M}_{1N\left(
0\right) }^{20} \\ 
\vdots & \ddots & \vdots \\ 
\mathcal{M}_{N-1,1\left( 0\right) }^{20} & \ldots & \mathcal{M}_{N-1,N\left(
0\right) }^{20}%
\end{array}%
\right) & \left( 
\begin{array}{ccc}
\mathcal{M}_{11\left( 0\right) }^{22}-\mathcal{M}_{NN\left( 0\right) }^{02}
& \ldots & \mathcal{M}_{1,N-1\left( 0\right) }^{22}-\mathcal{M}_{NN\left(
0\right) }^{02} \\ 
\vdots & \ddots & \vdots \\ 
\mathcal{M}_{N-1,1\left( 0\right) }^{22}-\mathcal{M}_{NN\left( 0\right)
}^{02} & \ldots & \mathcal{M}_{N-1,N-1\left( 0\right) }^{22}-\mathcal{M}%
_{NN\left( 0\right) }^{02}%
\end{array}%
\right)%
\end{array}%
\right) .
\]%
where

\[
\mathcal{M}_{ij}^{rn}=\mathcal{A}_{rn}^{\left( i0\right) }\delta _{ij}+%
\mathcal{C}_{rn}^{\left( ij0\right) },
\]%
and%
\begin{eqnarray*}
\mathcal{A}_{rn}^{\left( i0\right) } &=&\frac{g_{i}g_{j}}{J_{00}^{i}}%
\sum_{j=1}^{N}\int dK_{i}dK_{j}^{\prime }dP_{i}dP_{j}^{\prime }\gamma
_{ij}W_{\mathbf{pp}^{\prime }-\mathbf{kk}^{\prime }}^{ij}f_{i\mathbf{k}%
}^{\left( 0\right) }f_{j\mathbf{k}^{\prime }}^{\left( 0\right) }\left( E_{i%
\mathbf{k}}\right) ^{r}\left[ h_{ni}^{\left( 1\right) }\left( E_{\mathbf{p}%
i}-E_{\mathbf{k}i}\right) +h_{ni}^{\left( 2\right) }\left( E_{\mathbf{p}%
i}^{2}-E_{\mathbf{k}i}^{2}\right) \right] \\
\mathcal{C}_{rn}^{\left( ij0\right) } &=&\frac{g_{i}g_{j}}{J_{00}^{i}}\int
dK_{i}dK_{j}^{\prime }dP_{i}dP_{j}^{\prime }\gamma _{ij}W_{\mathbf{pp}%
^{\prime }-\mathbf{kk}^{\prime }}^{ij}f_{i\mathbf{k}}^{\left( 0\right) }f_{j%
\mathbf{k}^{\prime }}^{\left( 0\right) }\left( E_{i\mathbf{k}}\right)
^{r}\times \left[ h_{nj}^{\left( 1\right) }\left( E_{\mathbf{p}^{\prime
}j}-E_{\mathbf{k}^{\prime }j}\right) +h_{nj}^{\left( 2\right) }\left( E_{%
\mathbf{p}^{\prime }j}^{2}-E_{\mathbf{k}^{\prime }j}^{2}\right) \right]
\end{eqnarray*}%
with $W_{\mathbf{pp}^{\prime }-\mathbf{kk}^{\prime }}^{ij}=s\sigma
_{ij}\left( 2\pi \right) ^{6}\delta ^{4}\left( p_{i}+p_{j}^{\prime
}-k_{i}-k_{j}^{\prime }\right) $ being the transition rate, $N$ is the
number of hadrons considered, $\gamma _{ij}=1-\left( 1/2\right) \delta _{ij}$%
, $s$ is the Mandelstan variable, and we defined, 
\begin{eqnarray*}
h_{0i}^{\left( 0\right) }
&=&1+a_{10}^{(0)i}a_{10}^{(0)i}+a_{20}^{(0)i}a_{20}^{(0)i}, \\
h_{1i}^{\left( 0\right) }
&=&a_{11}^{(0)i}a_{10}^{(0)i}+a_{20}^{(0)i}a_{21}^{(0)i}, \\
h_{2i}^{\left( 0\right) } &=&a_{20}^{(0)i}a_{22}^{(0)i}, \\
h_{0i}^{\left( 1\right) }
&=&a_{10}^{(0)i}a_{11}^{(0)i}+a_{20}^{(0)i}a_{21}^{(0)i}, \\
h_{1i}^{\left( 1\right) }
&=&a_{11}^{(0)i}a_{11}^{(0)i}+a_{21}^{(0)i}a_{21}^{(0)i}, \\
h_{2i}^{\left( 1\right) } &=&a_{21}^{(0)i}a_{22}^{(0)i}, \\
h_{0i}^{\left( 2\right) } &=&a_{20}^{(0)i}a_{22}^{(0)i}, \\
h_{1i}^{\left( 2\right) } &=&a_{21}^{(0)i}a_{22}^{(0)i}, \\
h_{2i}^{\left( 2\right) } &=&a_{22}^{(0)i}a_{22}^{(0)i}.
\end{eqnarray*}

\section{Convergence of the Results with the Choice of $h$ and the Number of
SPH Particles}

\label{sec:convh}

\begin{figure}[ht]
\centering
\begin{tabular}{cc}
\includegraphics[width=0.3\textwidth]{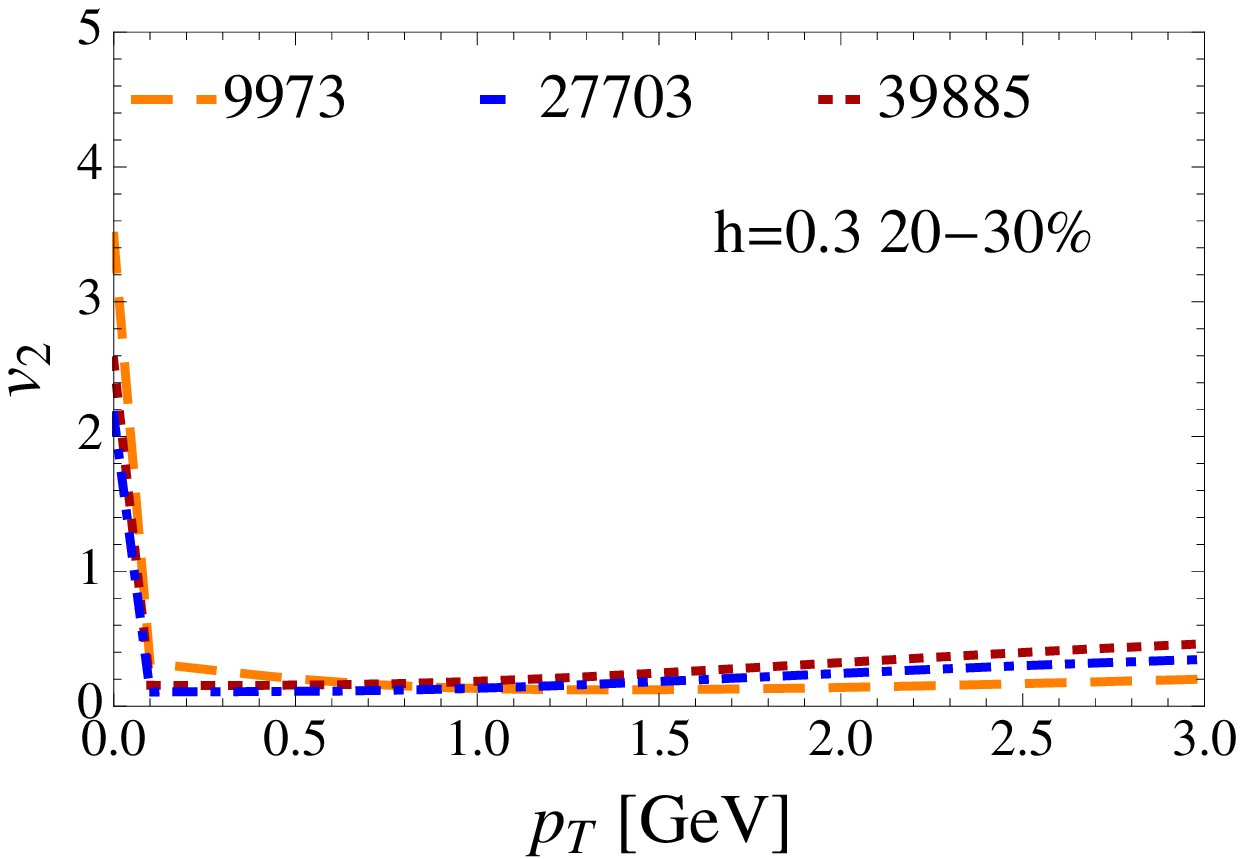} & %
\includegraphics[width=0.3\textwidth]{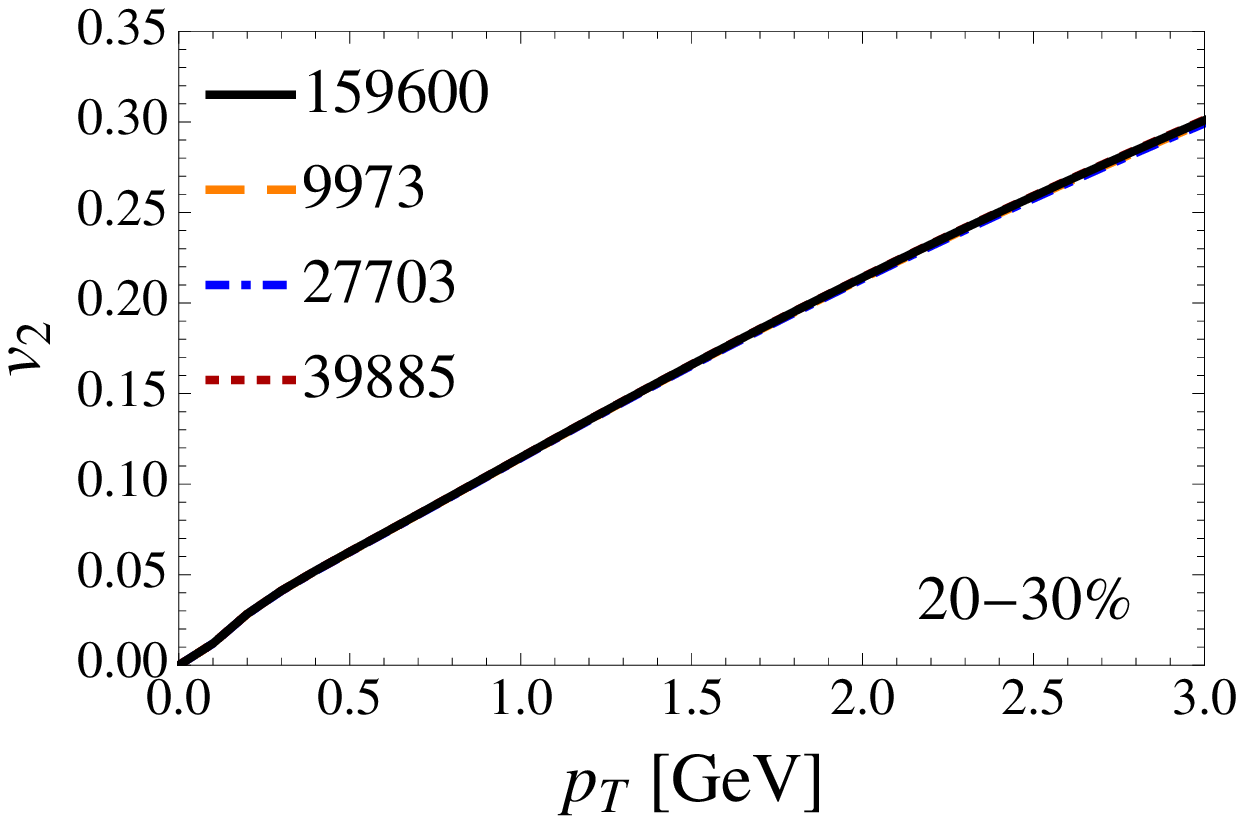} \\ 
\newline
\includegraphics[width=0.3\textwidth]{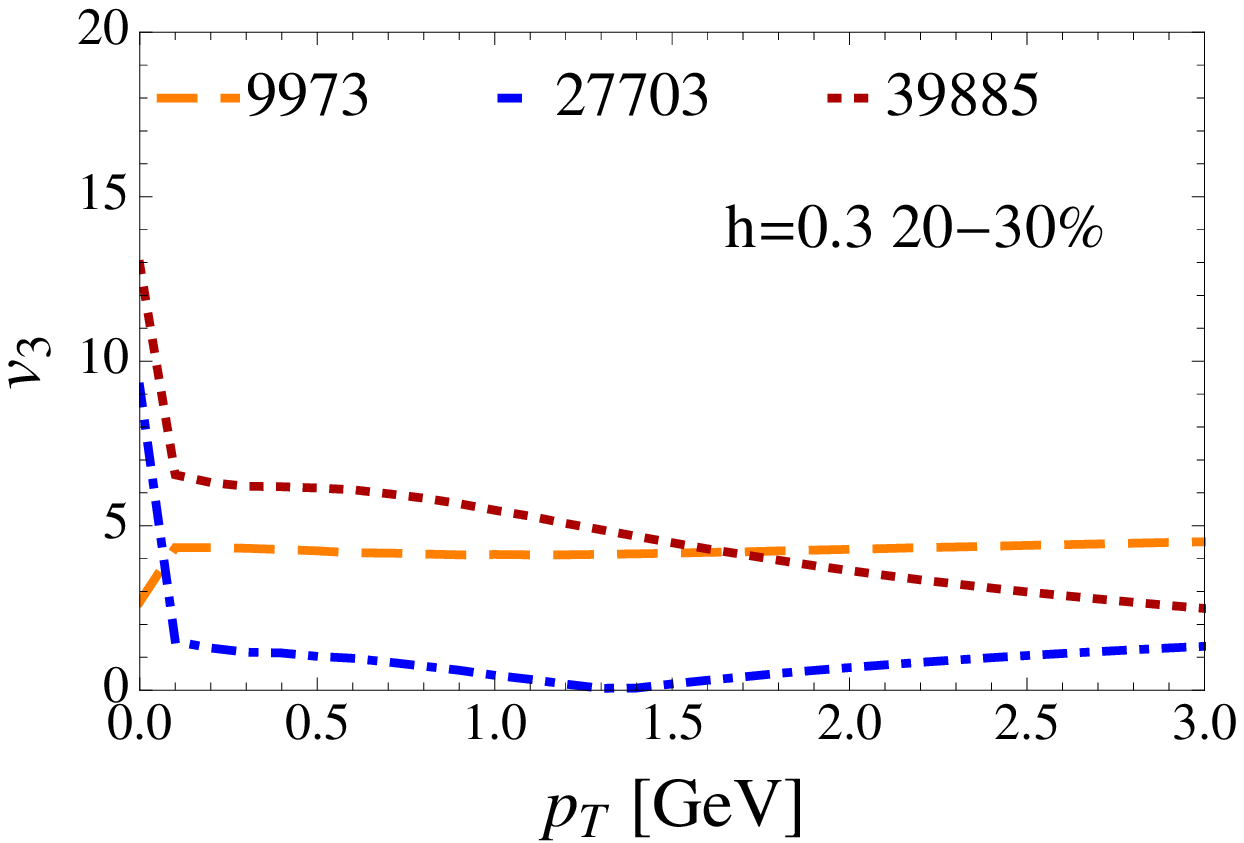} & %
\includegraphics[width=0.3\textwidth]{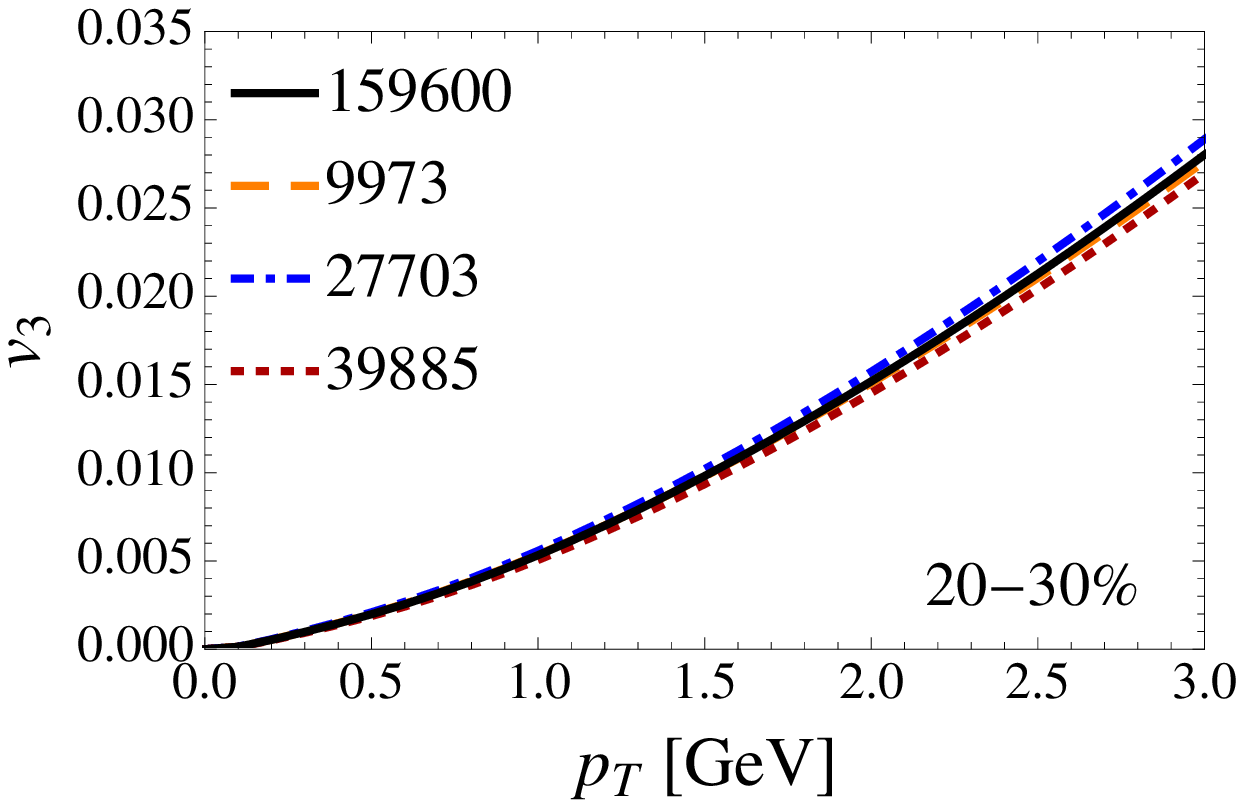} \\ 
\newline
\includegraphics[width=0.3\textwidth]{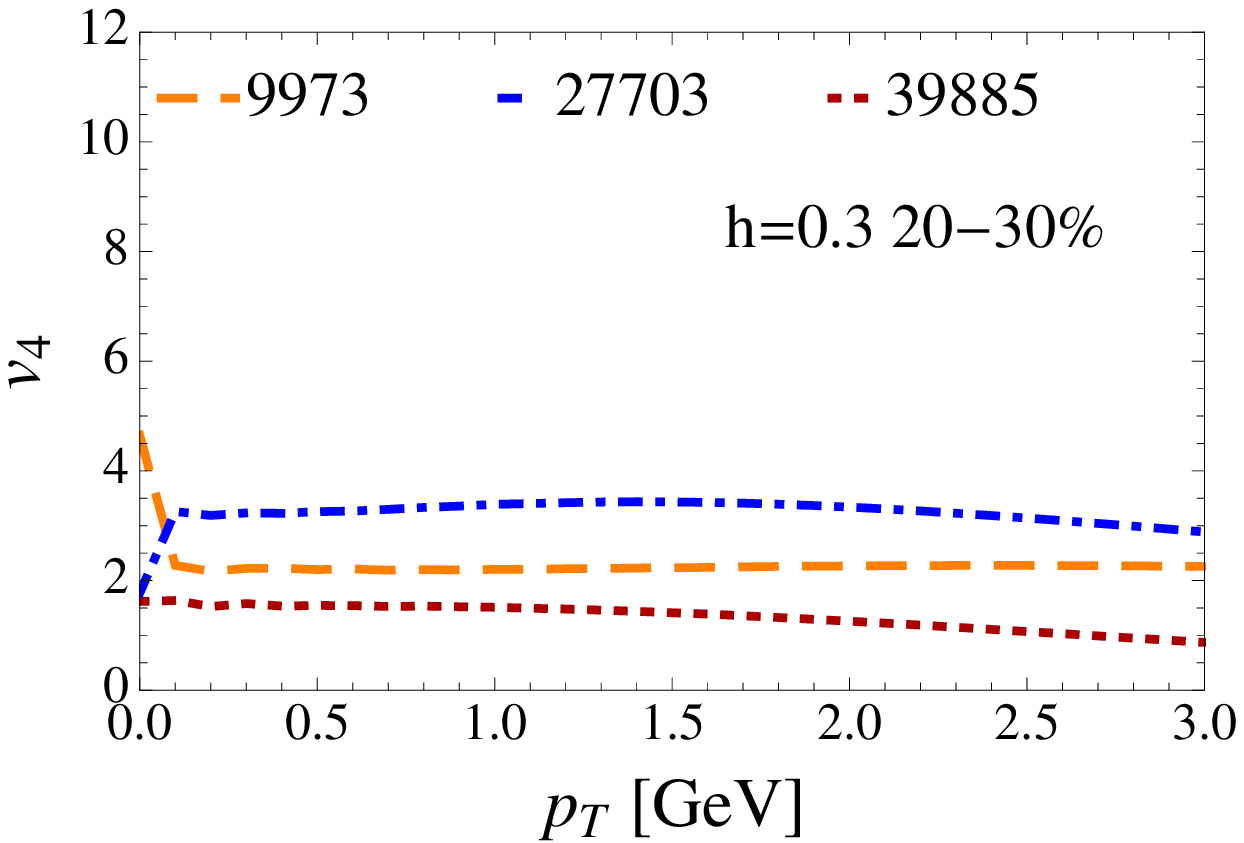} & %
\includegraphics[width=0.3\textwidth]{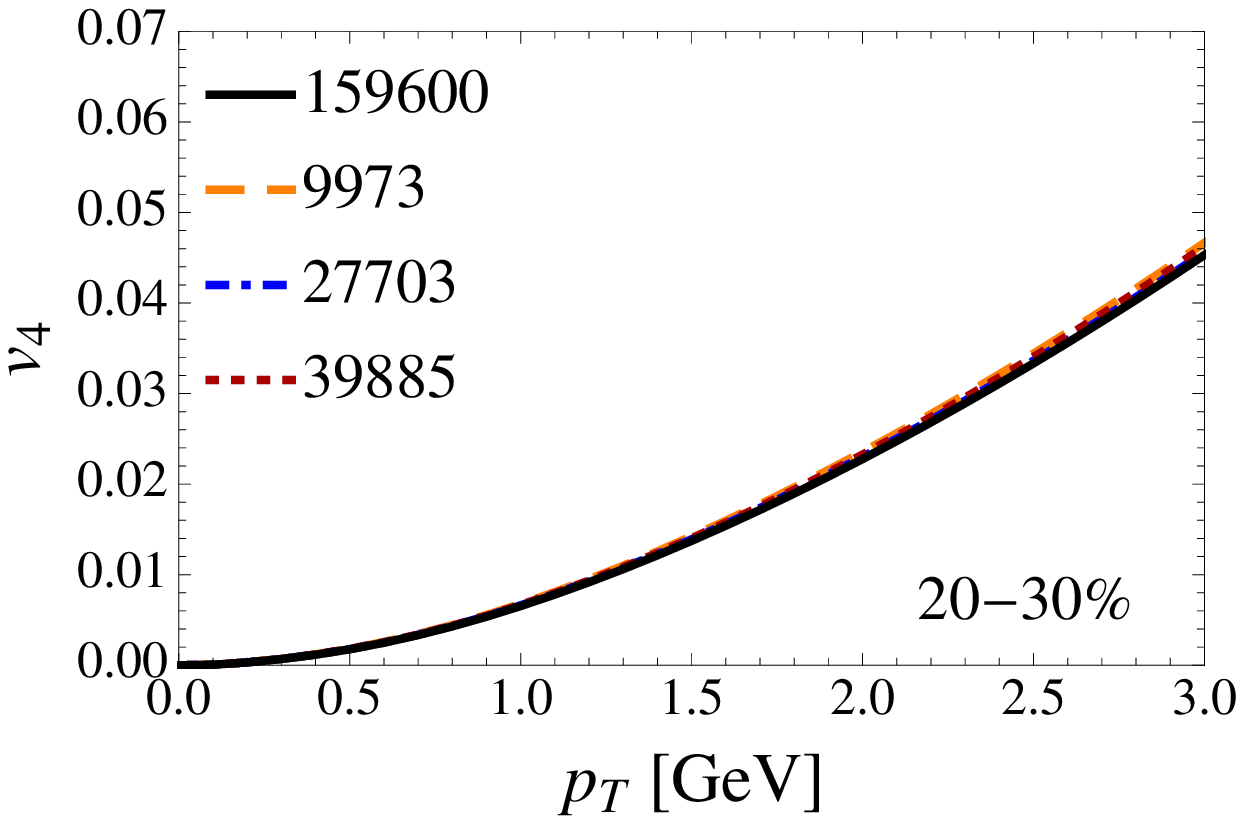} \\ 
\newline
\includegraphics[width=0.3\textwidth]{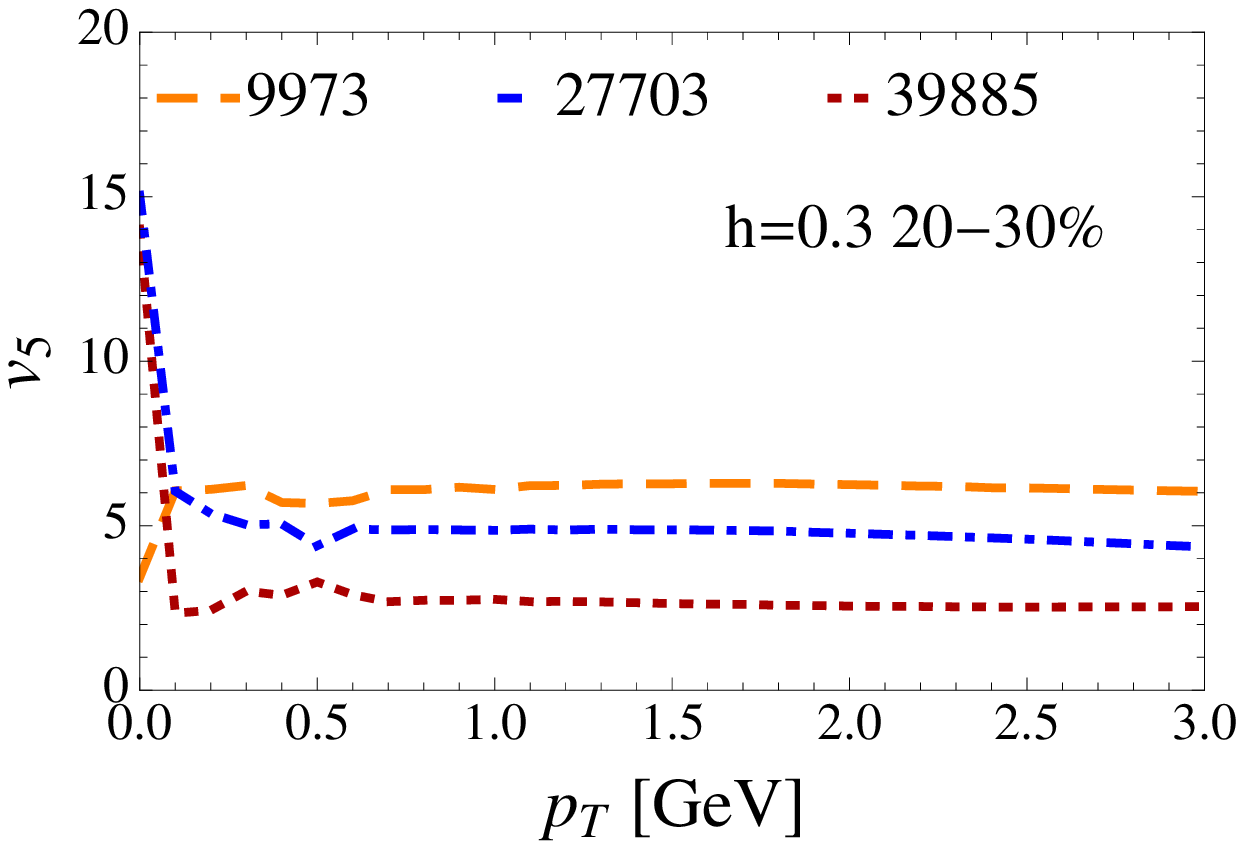} & %
\includegraphics[width=0.3\textwidth]{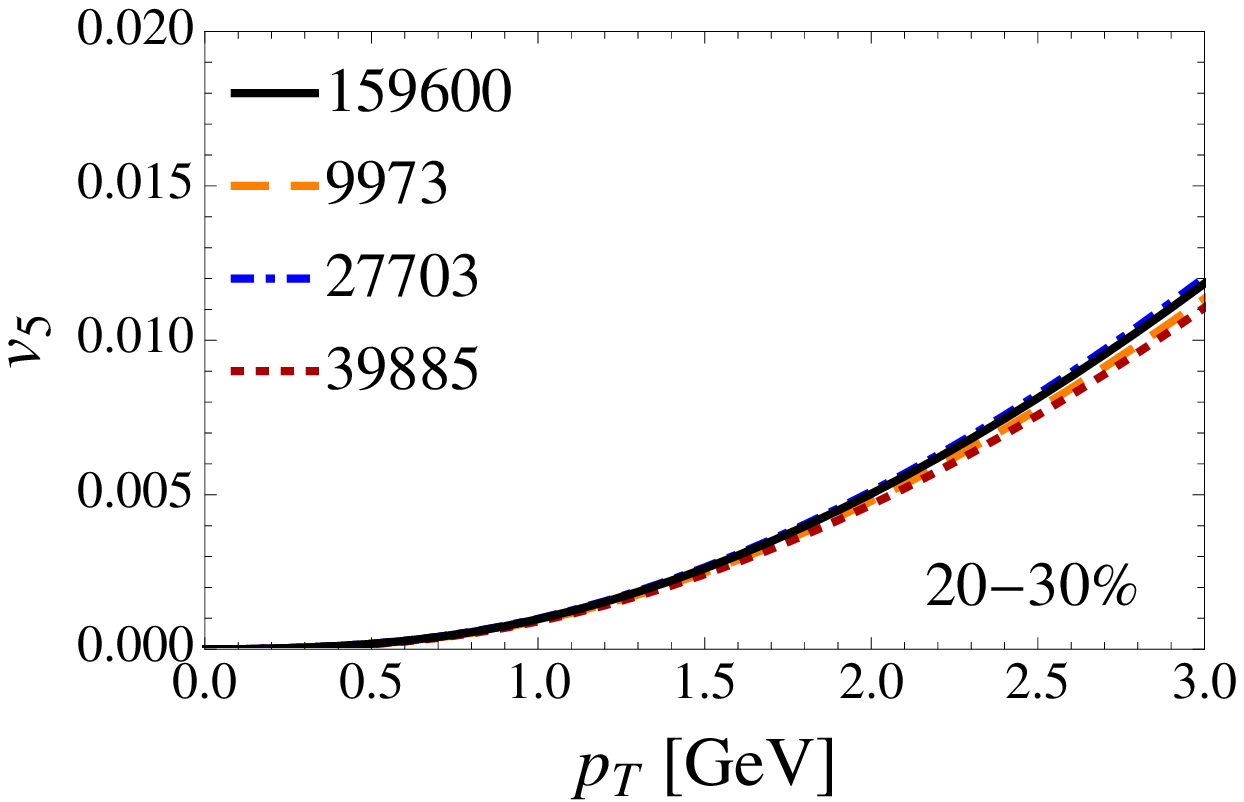} \\ 
\newline
\includegraphics[width=0.3\textwidth]{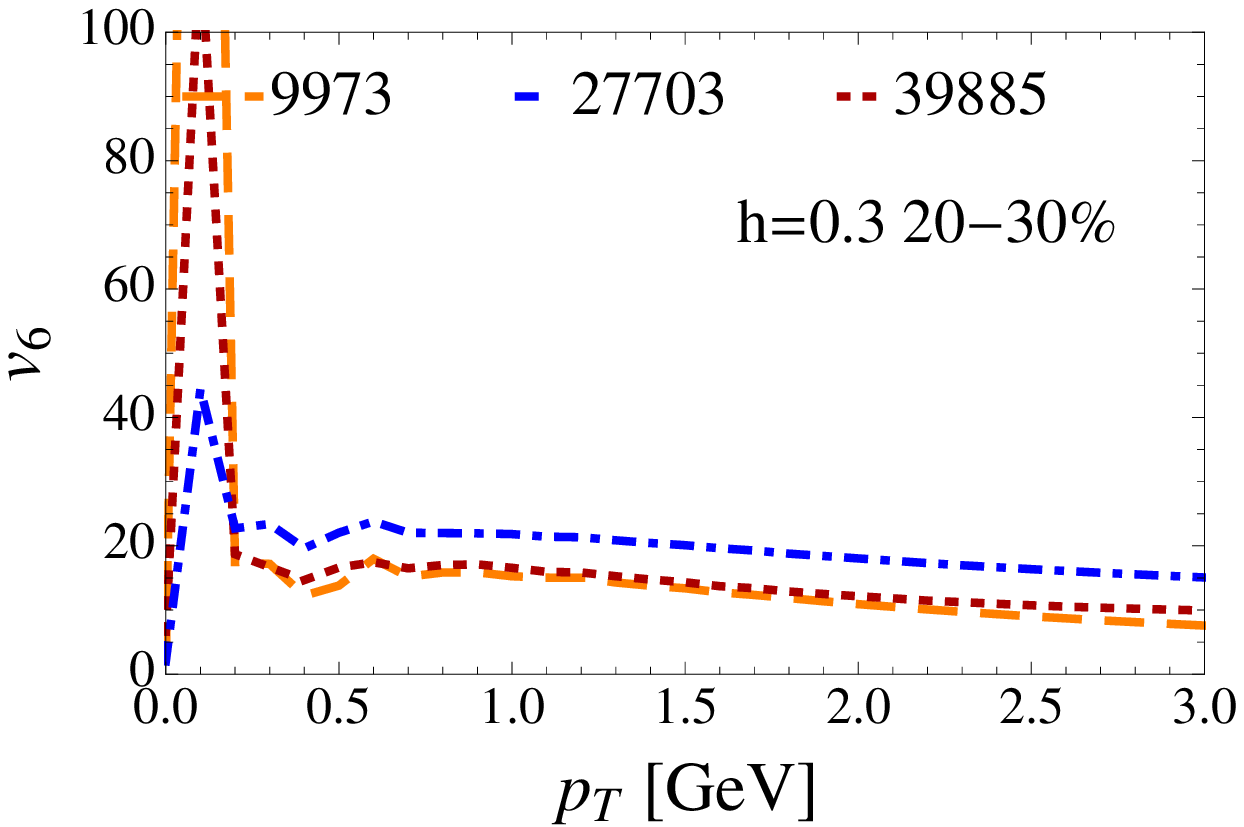} & %
\includegraphics[width=0.3\textwidth]{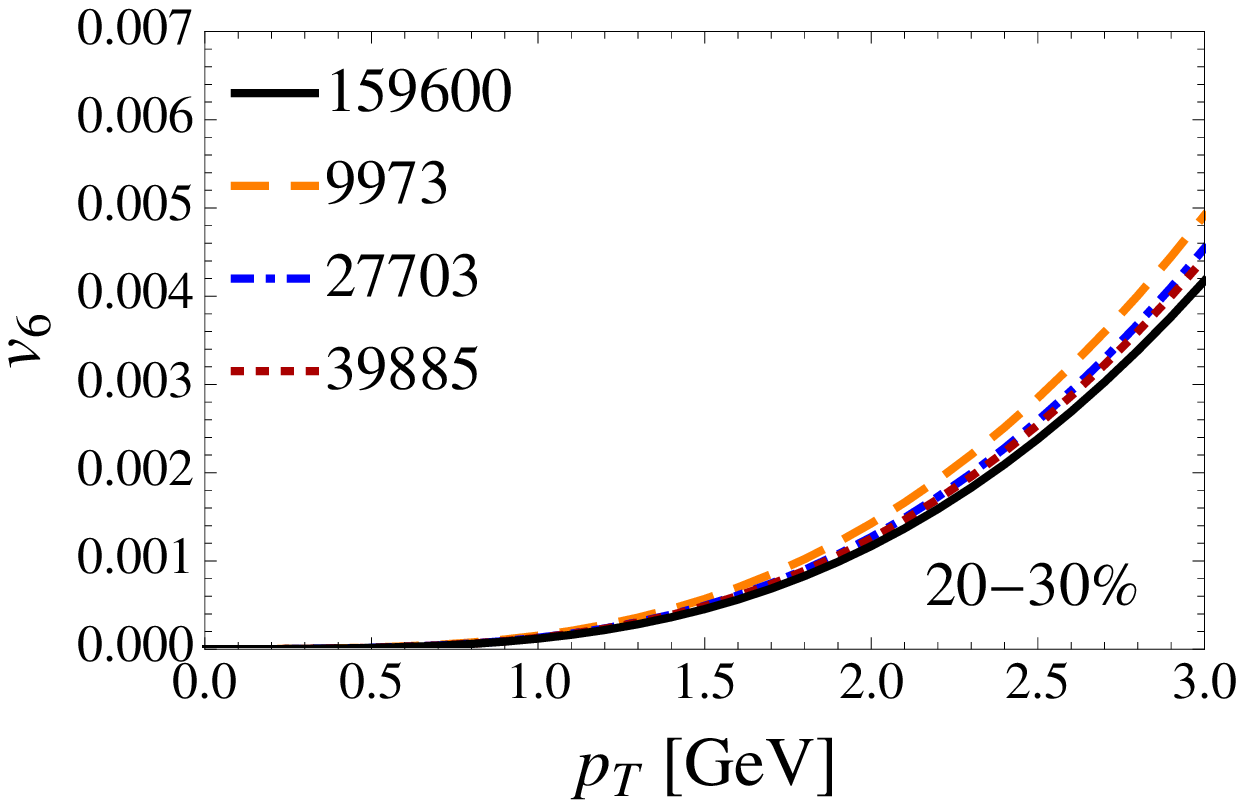} \\ 
& 
\end{tabular}%
\caption{Study of the convergence of the $v_n(p_T)$ coefficients with the
number of SPH particles for a fixed $h=0.3$ fm. On the left we show the
percentage deviation, see Eq.\ (\protect\ref{eqn:dev}), with respect to our
maximum of 159600 SPH particles. On the right we show the actual value of
the $v_n(p_T)$'s for the different $N_{SPH}$ values. We used a single
optical Glauber initial condition averaged over 150 events in the 20-30\%
centrality class.}
\label{tab:hconst}
\end{figure}

In this paper, the effects of bulk viscosity on collective flow coefficients
were computed using v-USPhydro with $h=0.3$ fm and a total number of SPH
particles ($N_{SPH}$) of roughly 3$\times10^4$. However, it is important to
test the convergence of our results with the choice of these parameters,
i.e, if one were to include more SPH particles would there be a change in
the flow harmonics and if so by what percentage? To demonstrate this we
first set $h=0.3$ fm and vary the number of SPH particles using averaged
initial conditions in the $20-30\%$ centrality bin. The initial condition
was averaged over roughly 150 events, which means that it is mostly smooth
but there is still some small remaining structure that generates small,
though nonzero, higher order flow harmonics. In Fig.\ \ref{tab:hconst} we
show our results for both the $v_n(p_T)$'s and the relative percentage
difference defined as 
\begin{eqnarray}  \label{eqn:dev}
v_n \%\left(N_{SPH}\right)=100\Big|\frac{v_n(N_{SPH})-v_n(N_{\infty})}{%
v_n(N_{\infty})}\Big|
\end{eqnarray}
where $N_{\infty}= 159600$ SPH particles is the maximum number of particles
set by our computational limitations.

As one can see in Fig.\ \ref{tab:hconst}, our choice of about 27000 SPH
appears reasonable with the corresponding $h=0.3$ fm (at very low $p_T$
there is a greater deviation because the $v_n$'s approach zero, which makes
the comparison more difficult). Quite generally, the different choices for $%
N_{SPH}$ differ from the ``infinite" $N_{\infty}$ limit by only $2-5\%$
depending on the specific $v_n(p_T)$ and the difference is practically
indiscernible with the naked eye (see the right plot in Fig.\ \ref%
{tab:hconst}). Elliptic flow is found to be by far the most robust. The one
exception being $v_6(p_T)$, which after this study we have decided not to
include in the paper in the results section due to the large deviation. To
perform a reliable study of $v_6(p_T)$ an extremely large number of SPH
particles would be needed, which would significantly slow down computation
time.

\begin{figure}[ht]
\centering
\begin{tabular}{cc}
\includegraphics[width=0.3\textwidth]{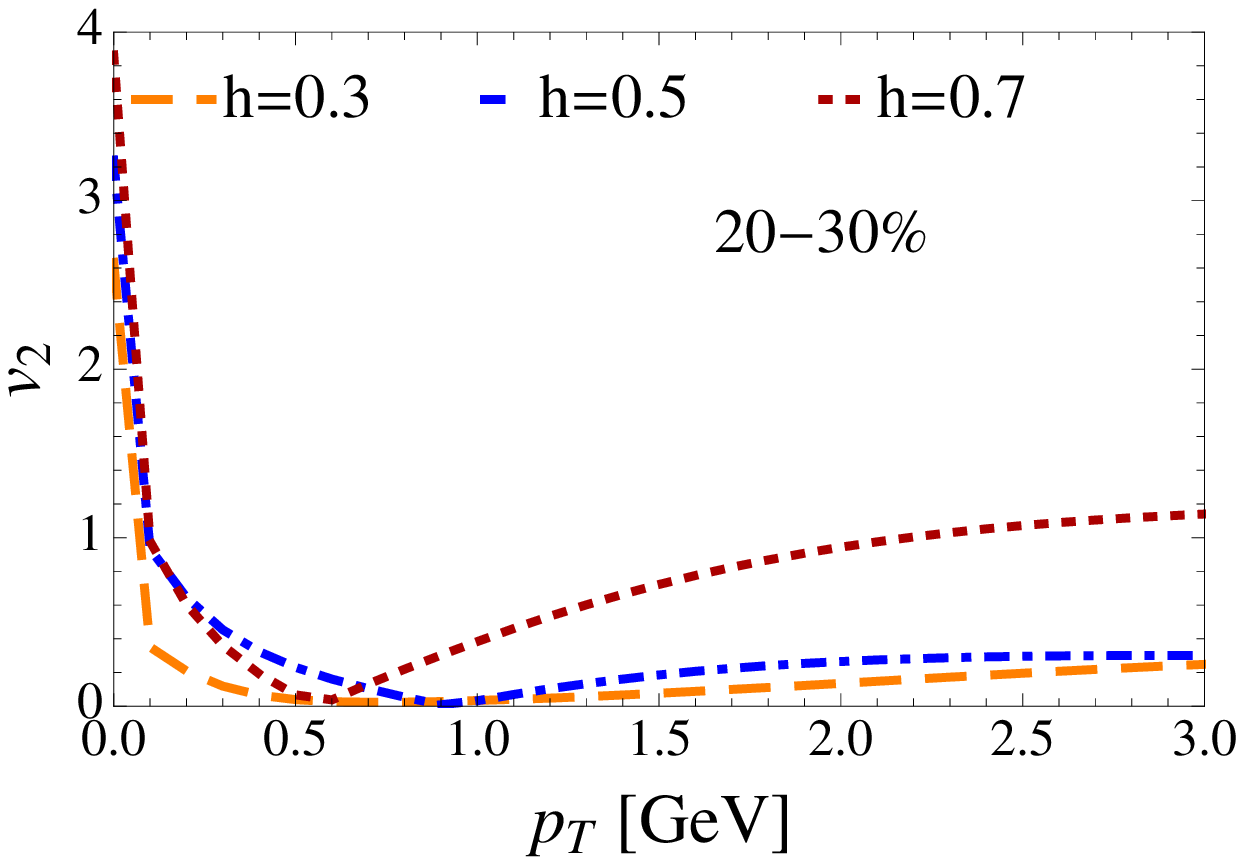} & %
\includegraphics[width=0.3\textwidth]{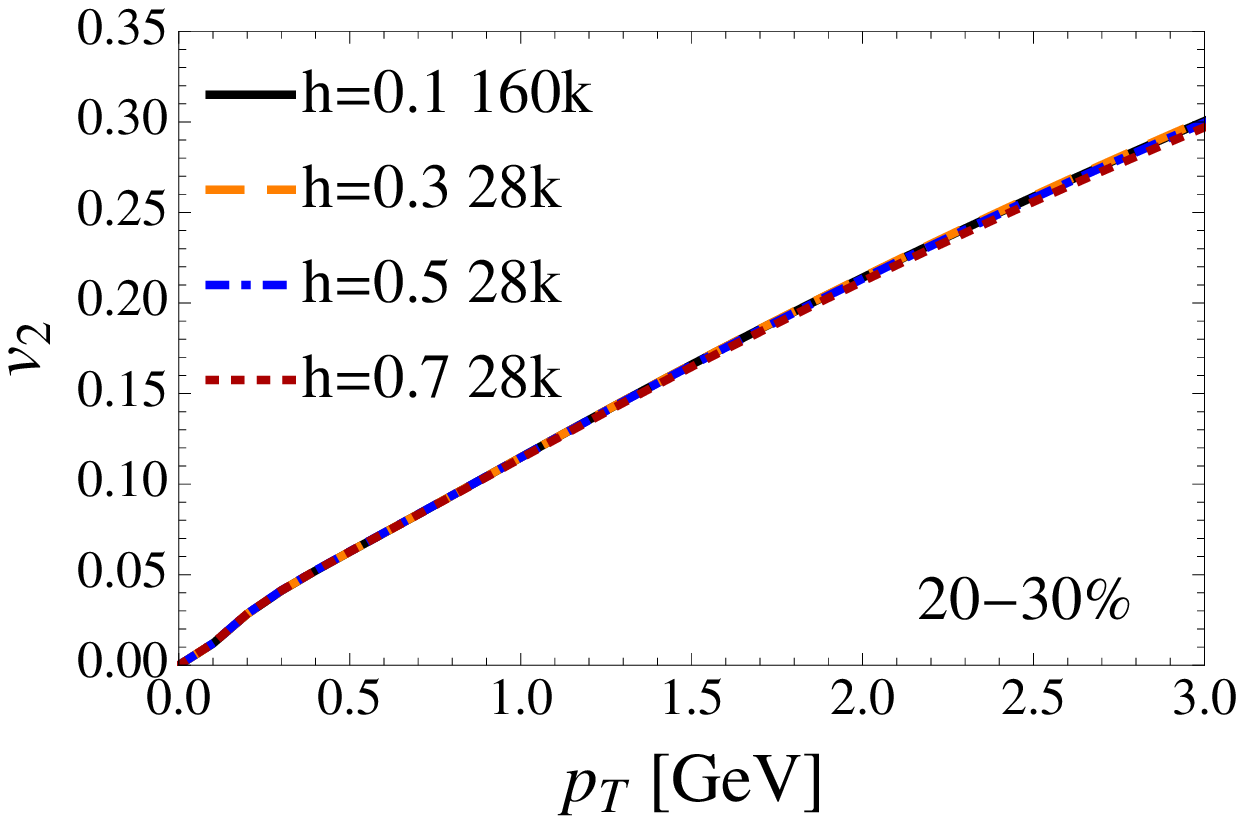} \\ 
\newline
\includegraphics[width=0.3\textwidth]{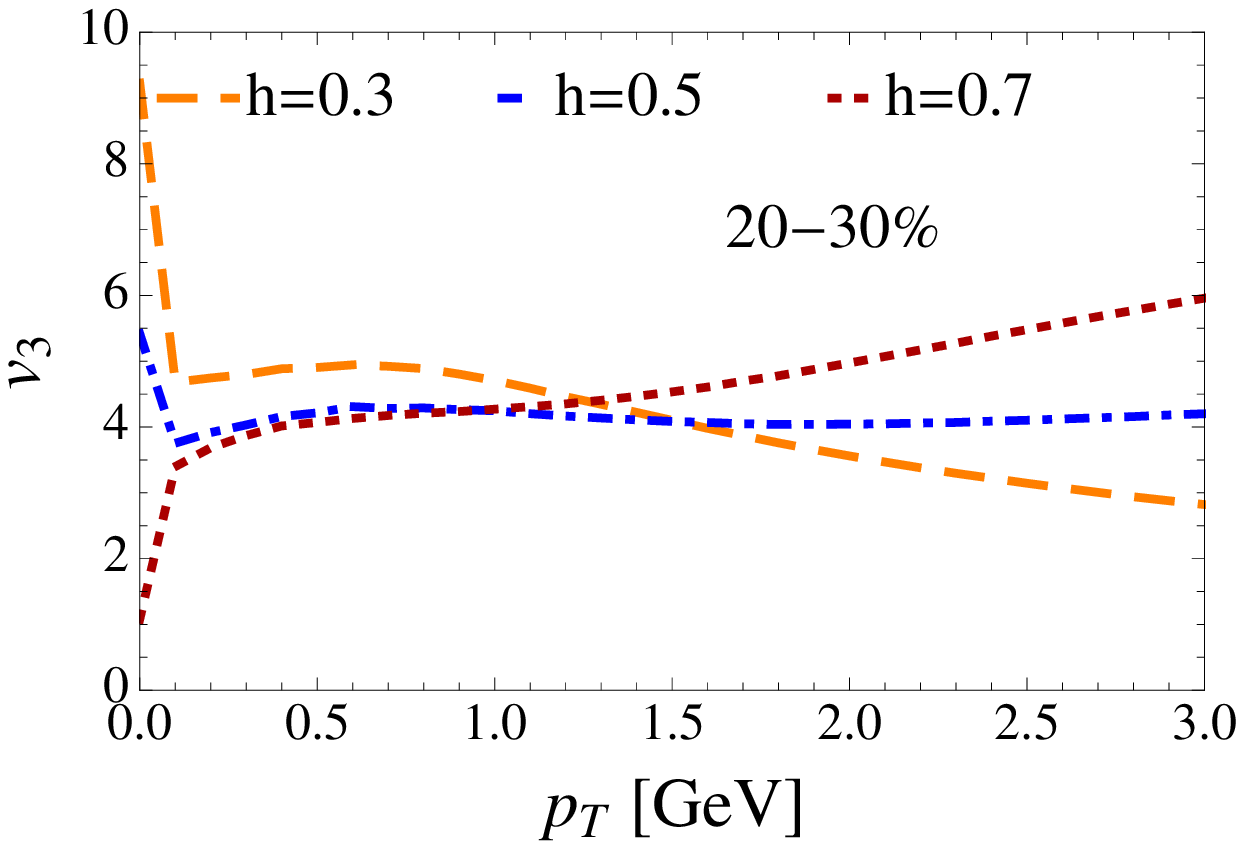} & %
\includegraphics[width=0.3\textwidth]{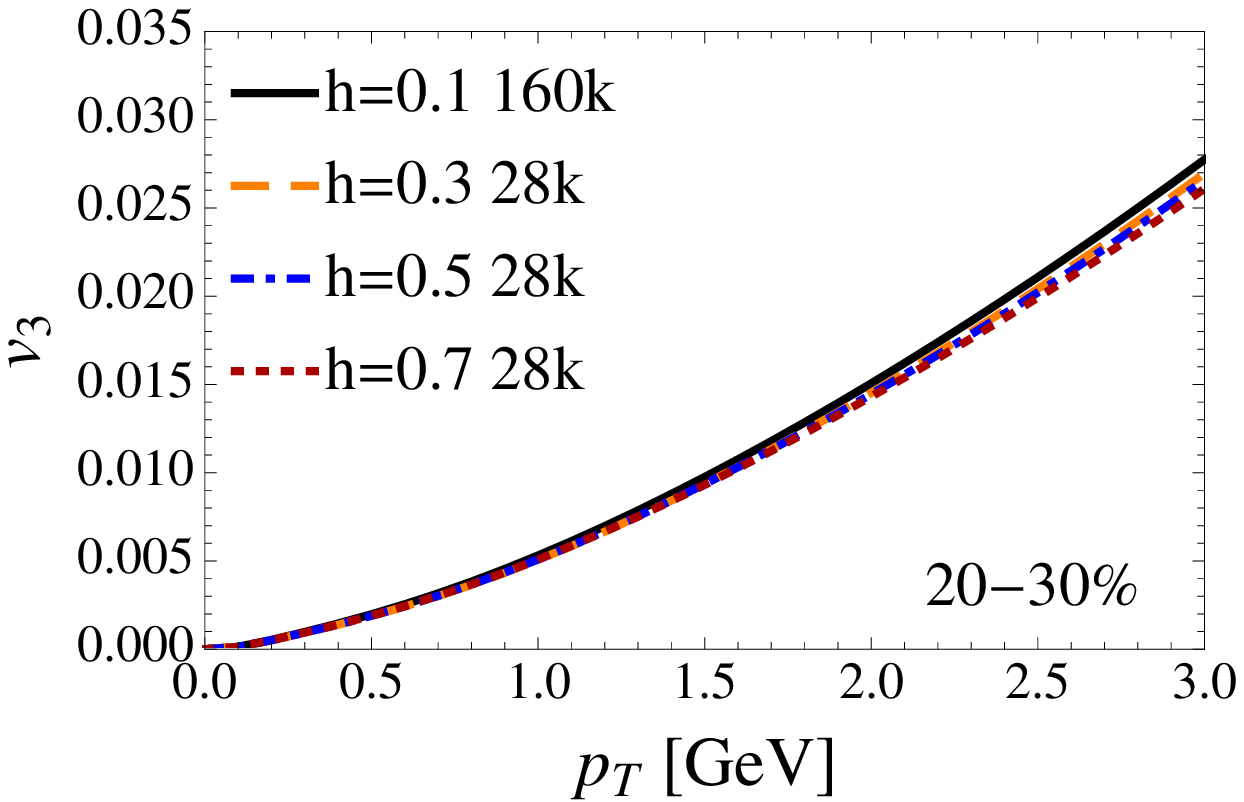} \\ 
\newline
\includegraphics[width=0.3\textwidth]{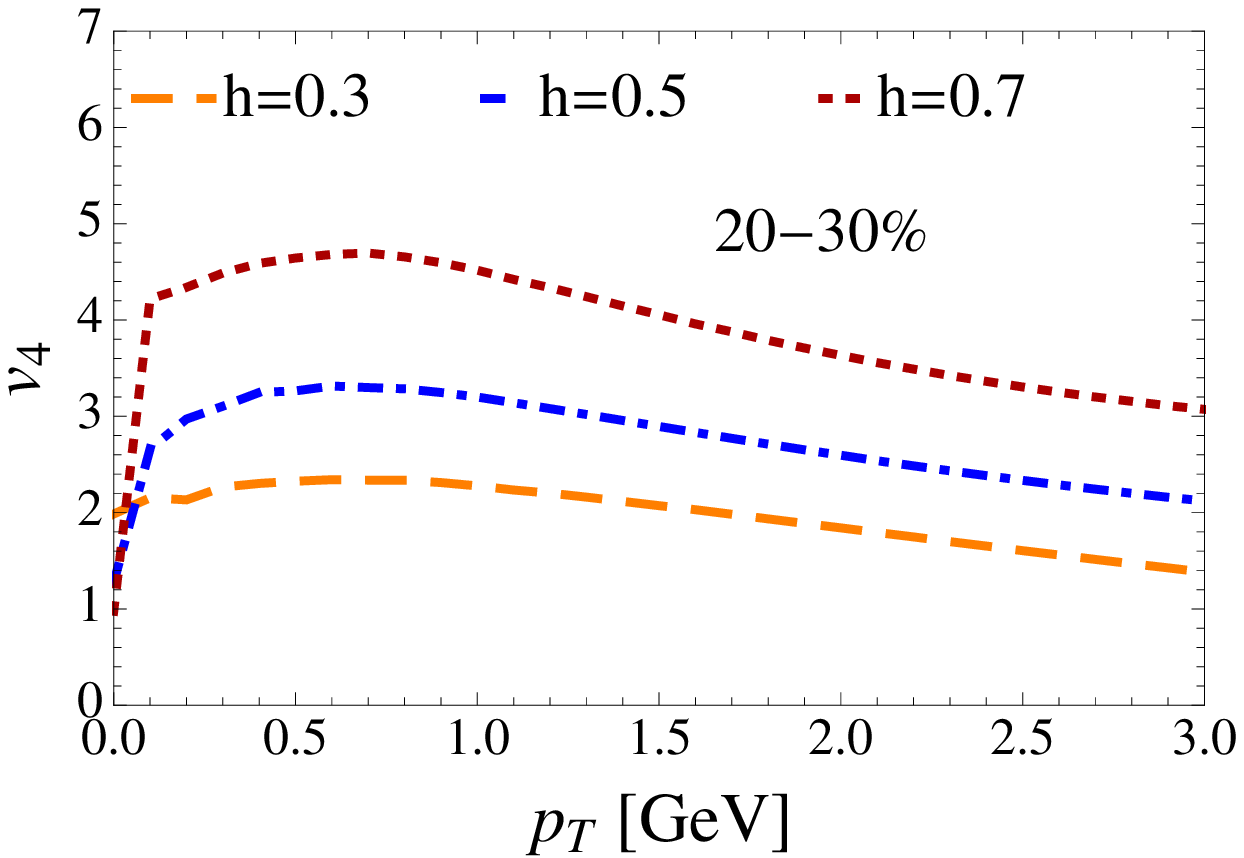} & %
\includegraphics[width=0.3\textwidth]{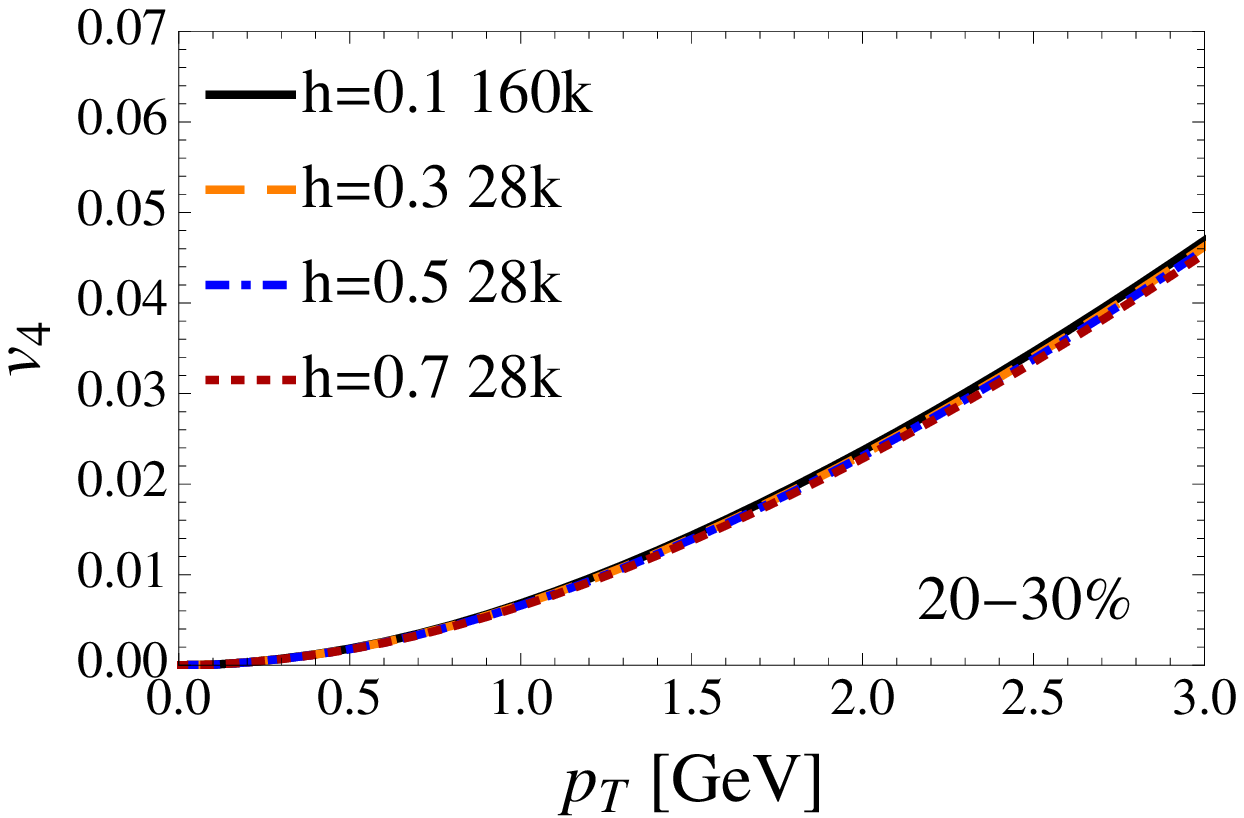} \\ 
\newline
\includegraphics[width=0.3\textwidth]{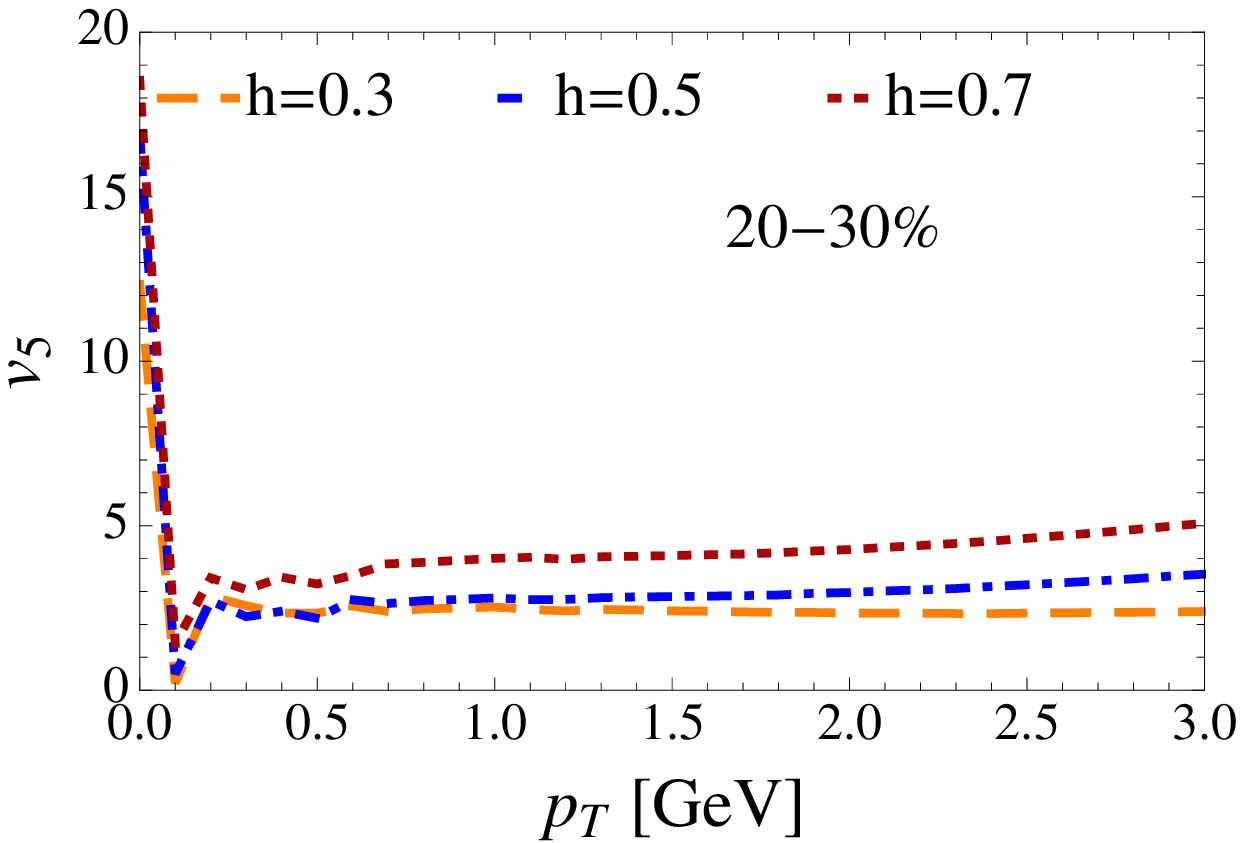} & %
\includegraphics[width=0.3\textwidth]{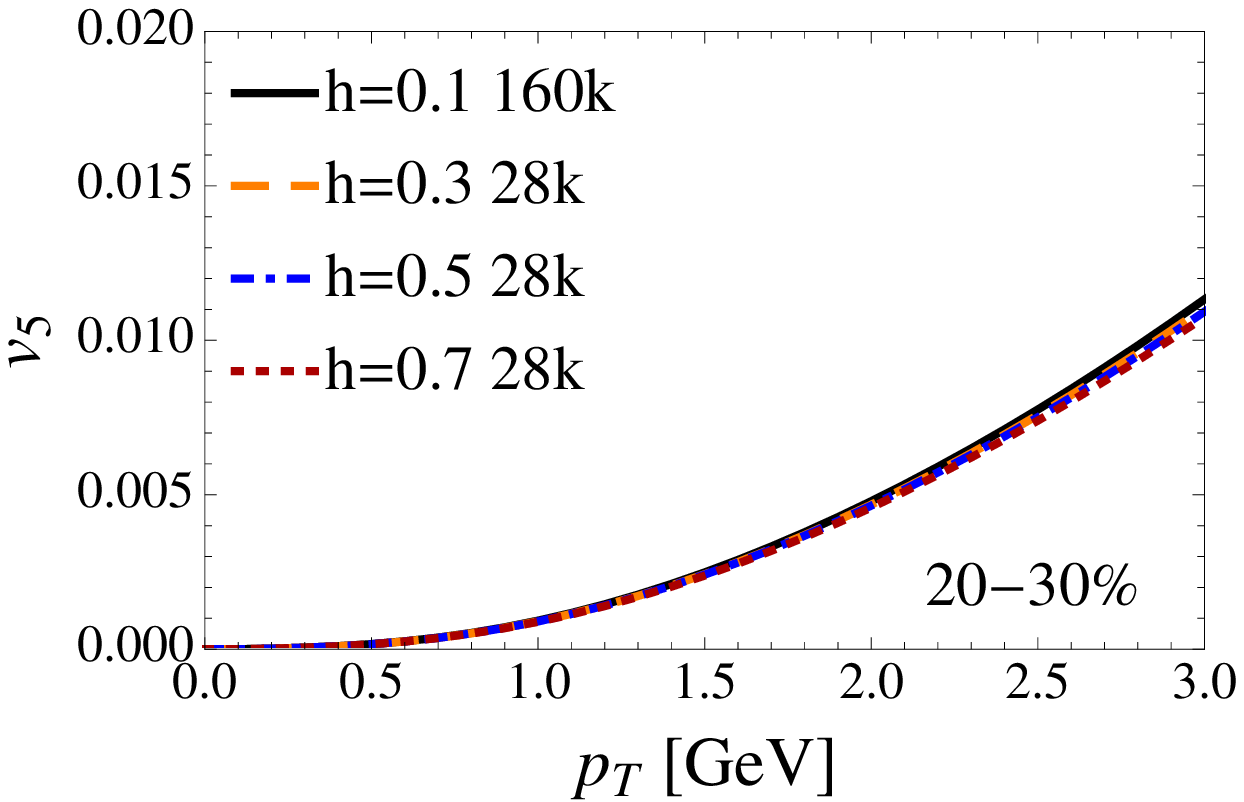} \\ 
\newline
\includegraphics[width=0.3\textwidth]{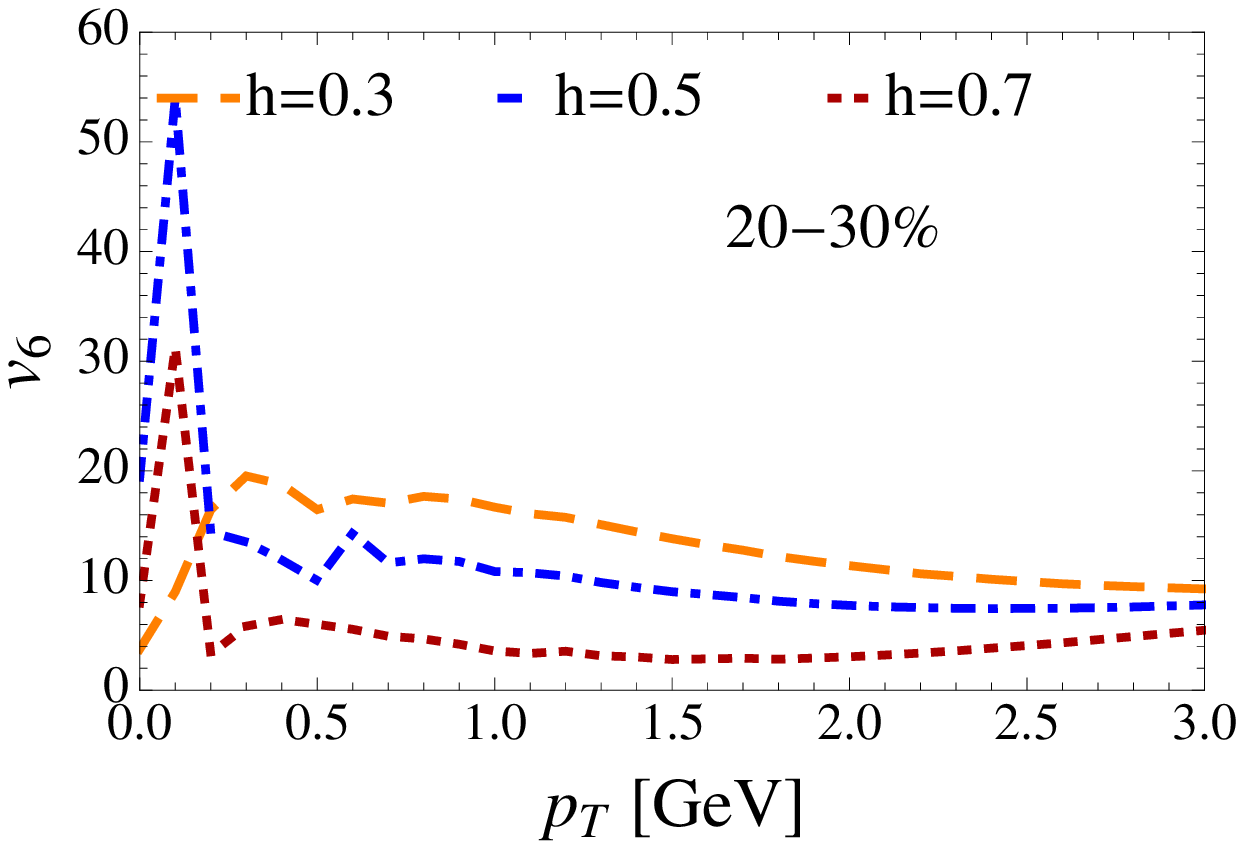} & %
\includegraphics[width=0.3\textwidth]{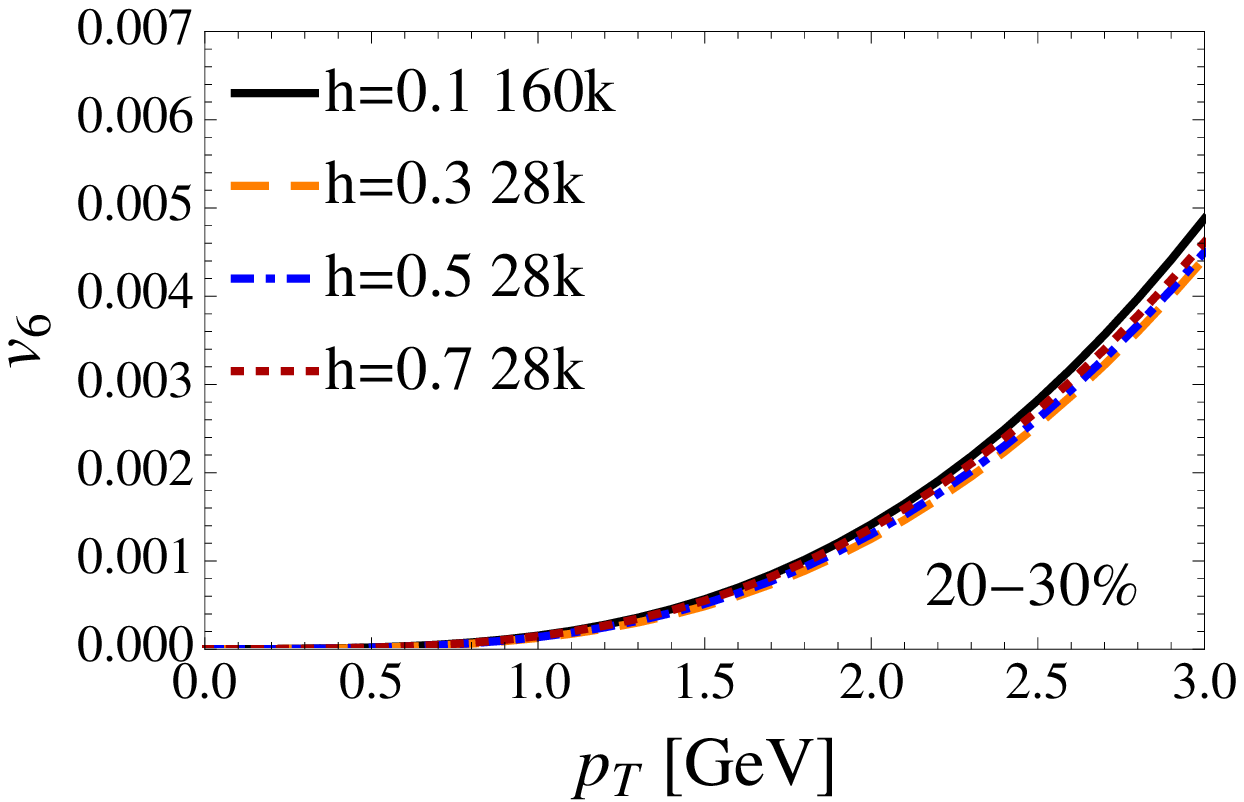} \\ 
& 
\end{tabular}%
\caption{The effect of the choice of $h$ on the total convergence of the $v_n
$'s ($h=0.3$ fm, $h=0.5$ fm, and $h=0.7$ fm). On the left the percentage
deviation (see Eq.\ (\protect\ref{eqn:dev}) is shown compared to $h=0.1$ fm
for 159,600 SPH particles. On the right we show the $v_n(p_T)$'s varying $%
N_{SPH}$ and $h$. We used a single optical Glauber initial condition
averaged over 150 events in the 20-30\% centrality class.}
\label{tab:hvary}
\end{figure}

Additionally, we test the convergence of our results with our choice for $h$%
. Because $h$ is a smoothing parameter, if we choose $h=0.3$ fm this
inherently limits our ability to probe very short length scales. Thus, it is
important that $h$ is small enough to take into account necessary
fluctuations but also large enough to allow for a reasonable computational
time. We show the variation of our results for the flow coefficients with
the choice of $h$ in Fig.\ \ref{tab:hvary}. The left plot shows the
percentage difference with respect to the $N_{\infty}$ limit and $h=0.1$ fm
while the right plot shows the actual values of the coefficients versus $p_T$
for various values of $h$ and $N_{SPH}$. One can see in Fig.\ \ref{tab:hvary}
that the difference between $h=0.5$ fm and $h=0.3$ fm is not that large,
however, results for $h=0.7$ fm consistently show a larger deviation for $v_2
$ to $v_5$. This suggests that for the averaged Glauber initial conditions
used in these tests most of the important structure is larger than $0.5$ fm.
Clearly, if one were to consider other types of initial conditions which
display structure at smaller length scales, such as those that include gluon
saturation effects \cite{Gale:2012rq}, the type of analysis discussed in
this Appendix must be performed again.

\end{document}